\documentclass[useAMS]{mn2e}

\usepackage{amssymb,amsmath,psfig,times}
\def\gsim{ \lower .75ex \hbox{$\sim$} \llap{\raise .27ex \hbox{$>$}} }
\def\lsim{ \lower .75ex\hbox{$\sim$} \llap{\raise .27ex \hbox{$<$}} }

\def\sc{Schwarzschild}
\def\beq{\begin{equation}}
\def\eeq{\end{equation}}

\def\sc{Schwarzschild}

\title[BLR and $\gamma$--ray luminosities in blazars]
{The relation between broad lines and $\gamma$--ray luminosities 
in {\it Fermi} blazars}
\author[T. Sbarrato et al.]
{T. Sbarrato$^1$\thanks{Email: tullia.sbarrato@brera.inaf.it}, 
G. Ghisellini$^2$, 
L. Maraschi$^3$,
M. Colpi$^1$.
\\
$^1$Department of Physics G.\ Occhialini, University of Milano-Bicocca, 
Piazza della Scienza 3, 20126 Milano, Italy\\
$^2$INAF -- Osservatorio Astronomico di Brera, Via Bianchi 46, I--23807 Merate, Italy\\
$^3$INAF -- Osservatorio Astronomico di Brera, Via Brera 28, I--20100 Milano, Italy\\
}

\begin{document}  

\maketitle

\begin{abstract}
We study the relation between the mass accretion rate, the jet power, and the black hole
mass of blazars.
To this aim, we make use of the Sloan Digital Sky Survey (SDSS) and the
11 months catalog of blazars detected at energies larger than 100 MeV by the 
Large Area Telescope (LAT) onboard the {\it Fermi} satellite.
This allows to construct a relatively large sample of blazars
with information about the luminosity (or upper limits) of their emission lines
used as a proxy for the strength of the disc luminosity and on the luminosity
of the high energy emission, used as a proxy for the jet power.
We find a good correlation between the luminosity of the broad lines and the 
$\gamma$--ray luminosity as detected by {\it Fermi}, both using absolute values
of the luminosities and normalising them to the Eddington value.
The data we have analyzed 
confirm that the division of blazars into BL Lacs and Flat Spectrum Radio 
Quasars (FSRQs) is controlled by the line luminosity in Eddington units.
For small values of this ratio 
the object is a BL Lac,
while it is a FSRQs for large values.
The transition appears smooth, but a much larger number of objects is needed to
confirm this point.
\end{abstract}

\section{Introduction}

The classic division between Flat Spectrum Radio Quasars (FSRQs) and
BL Lac objects is 
mainly
based on the Equivalent Width (EW) of the emission lines.
Objects with rest frame EW$>5$ \AA\ are classified as FSRQs 
(see e.g. Urry \& Padovani 1995).
Marcha et al. (1996) and Landt, Padovani \& Giommi (2002)
discussed the Ca H\&K 4000--\AA\ (rest frame)
break as a criterion helping to distinguish BL Lac objects from
low--luminosity radio galaxies.
Furthermore, Marcha et al. (1996) proposed that objects with a weak Ca break and
with EW even larger than 5 \AA\ should be classified
as BL Lacs. 
On the other hand, Scarpa \& Falomo (1997) showed a continuity 
between BL Lac and FSRQs concerning the luminosity of the Mg II line,
taken as an indication against a clear separation of blazars
in the two subclasses.
Landt et al. (2004), instead, considered {\it narrow} lines, 
such as [O II] and [O III], and found that it is possible to 
separate {\it intrinsically}
weak and strong line blazars in the [O II] and [O III] equivalent width
plane.

The classification scheme based on the EW of the broad lines 
has been adopted both because it is 
observationally simple, and because it was thought to measure the 
relative importance of the non--thermal jet emission over the thermal one.
However, we now know that the jet electromagnetic output is often dominated by 
the emission at higher energies (hard X--rays and $\gamma$--rays), and therefore
the EW of the optical emission lines is not a good measure of the jet dominance.
Furthermore, the jet flux is much more variable than the underlying thermal
emission, causing the measured EW to vary. 
Occasionally, a blazar
with very luminous emission lines, that should be classified as
a FSRQ, can instead appear as BL Lacs when the optical jet flux is particularly strong.
Conversely, 
a BL Lac  
in a particular faint state could show broad emission lines that, albeit weak,
can have EW greater than 5 \AA.
In Ghisellini et al. (2011; G11 thereafter) we have therefore proposed a more physical distinction
between the two classes of blazars, based on the luminosity of the broad 
emission lines measured in Eddington units: $L_{\rm BLR}/L_{\rm Edd}$.
We proposed that when $L_{\rm BLR}/L_{\rm Edd} \gsim 5\times 10^{-4}$ 
the objects are FSRQs, and are BL Lacs below this value. 
Normalizing to the Eddington luminosity ensures the appropriate comparison
among objects of different black hole masses.

The sample of blazars studied in G11 was limited, since it was  based
on a small sub--sample of bright FSRQs detected in $\gamma$--rays by the 
{\it Fermi} satellite during the first 3 months of operation 
(LBAS sample, Abdo et al. 2009), and on BL Lac objects detected 
by {\it Fermi} during the first 11 months (1LAC sample, Abdo et al. 2010a), 
with a relatively steep $\gamma$--ray energy spectral index $\alpha_\gamma$ 
($\alpha_\gamma>1.2$).
These BL Lacs occupy the region of the spectral index -- $\gamma$--ray luminosity 
($\alpha_\gamma$--$L_\gamma$) plane occupied mainly by FSRQs 
(see Ghisellini, Maraschi \& Tavecchio 2009).

Since the broad emission lines are produced by clouds photo--ionized by the 
radiation produced by the accretion disc, there is a direct relation 
between $L_{\rm BLR}$ and the accretion disc luminosity $L_{\rm d}$. 
Therefore, measuring the broad line luminosities, 
we have information on the disc luminosity even when it is not directly visible,
as often occurs in blazars whose optical continuum is dominated by the jet flux.
In turn, by knowing $L_{\rm BLR}$ and the bolometric jet luminosity,
we can then study the relation between the jet and the accretion power.
This in fact is the final aim of these studies.
Earlier attempts to find the ratio between the jet and the 
accretion power were done by e.g. Celotti, Padovani \& Ghisellini (1997) and 
by D'Elia, Padovani \& Landt (2003): the novelty here is on one hand the
way to estimate the jet power, and on the other hand  
the large number of sources for which the $\gamma$--ray detection 
ensures a good estimate of the jet power (or at least a good proxy for it),
coupled with the large number of blazars present in the Sloan Digital Sky Survey (SDSS;
York et al.\ 2000) with 
spectroscopic data.

Another important ingredient for this line of research is the black hole mass,
allowing to measure luminosities and powers in Eddington units.
Besides allowing to compare objects
with different black hole masses, it allows to investigate
if the accretion regime has indeed a transition, from radiatively
efficient to inefficient, when the mass accretion rate 
in Eddington units $\dot M/\dot M_{\rm Edd}$ goes below a critical value
(see e.g. Narayan, Garcia \& McClintock 1997), and to see how this influences the jet power.
For instance, the division between BL Lacs and FSRQs in the sample
of blazars detected during the first three months of the {\it Fermi} all sky survey (LBAS)
seems to corresponds to disc luminosities $L_{\rm d}/L_{\rm Edd}\sim 10^{-2}$
(Ghisellini, Maraschi \& Tavecchio 2009; see also the earlier proposal
about the division of FR I and FR II radio--galaxies in Ghisellini \& Celotti 2001).

For these reasons we are motivated to enlarge the original sample of G11,
studying all blazars for which we can have information about their emission lines
(as a proxy for the disc luminosity), 
their $\gamma$--ray luminosity (as a proxy for the jet power), and their black hole mass.
The two largest samples useful for this study are the SDSS 
and the {\it Fermi} 1LAC sample.
In \S 2 we present the samples used for this work and in \S 3 we discuss how
we have derived the broad line luminosities, or their upper limits.
In \S 4 we present the relation between $L_{\rm BLR}$ and the $\gamma$--ray luminosities,
and we discuss our findings in \S 5.

\section{The samples}

We are interested in grouping a large number of blazars 
with reliable measures of Broad Line Region and $\gamma$--ray luminosities.
The Sloan Digital Sky Survey (SDSS), that provides the largest publicly
available catalog of spectral objects,
and the Large Area Telescope (LAT) onboard the \textit{Fermi Gamma--Ray
Space Telescope} constitute optimal devices to investigate 
in this direction.

We tried to select for our analysis the largest group of
blazars with reliable redshift and black hole mass measures.
At first, we grouped a sample of optically selected quasars
from the SDSS (seventh data release, DR7), that have been 
deeply analysed by Shen et al. (2011; hereafter S11) and are {{\it Fermi}--detected.
Furthermore, in order to enlarge our analysis towards lower luminosities, 
we included in our study an optically selected group of BL Lac candidates.
The BL Lacs that we took in consideration are supposed to be lineless,
but their redshift and black hole masses have been derived in the 
work by Plotkin et al. (2011; hereafter P11) from the galaxy spectral absorption features.
In the end, we tried to look for possible intermediate objects,
that have been excluded both from SDSS DR7 quasar catalog and 
P11 BL Lacs catalog.
For this purpose, we selected a small group of AGNs from 
the previous SDSS data release 
(DR6, Adelman--McCarthy et al.\ 2008) that are \textit{Fermi}--detected.

\subsection{1LAC sample}

This is the AGN sample resulting after 11 months of
operation of the Large Area Telescope (LAT) onboard the {\it Fermi} satellite 
(First LAT AGN Catalog, 1LAC, Abdo et al. 2010a).
As revised in Ackermann et al. (2011), this sample is made by 671 sources 
at high Galactic latitude ($|b| > 10^\circ$).
The statistical significance of the {\it Fermi} detection
was required to be TS$>$25
(TS stands for Test Statistics, see Mattox et al. 1996 for the definition.
TS=25 approximately corresponds to 5$\sigma$).

By requiring that the associations of the detected sources
have a probability $P\geq80\%$ and that there is only one
AGN in the positional error box of the {\it Fermi} detection, the 
{\it Fermi} team has constructed the {\it clean} 1LAC sample,
composed of 599 AGNs.
Of these, 248 are classified by the {\it Fermi}/LAT team as 
FSRQs and 275 as BL Lacs. 
The remaining sources are either of unknown blazar type (50) or 
non--blazar AGNs (26).
We then focus on the  248 FSRQs and 275 BL Lacs.
All FSRQs have known redshift, while for about half of the BL Lac
objects the redshift is still unknown.

\subsection{SDSS DR7 Quasar sample}

We have collected at first a group of optically selected quasars.
The SDSS provides a quasar catalog out of the Data Release 7, 
that includes 105783 objects and has been spectrally analysed by S11.
This sample includes quasars that have luminosities larger than $M_i=-22$ 
[i.e.\ $\nu L_{\nu}(5100$ \AA$) = 10^{44}$ erg s$^{-1}$], that have at least 
one emission line with FWHM$>1000$ km s$^{-1}$ and a reliable spectroscopical 
redshift. For this group of sources S11 calculated continuum and emission line measurements 
around the H$\alpha$, H$\beta$, MgII and CIV regions, and derived 
virial black hole masses with different calibrations,
along with what they consider the best estimate.
The broad line luminosities and the best estimate of the black hole masses
are considered for our work.

From the cross--correlation with the 1LAC sample, we obtained a group of 
49 {\it Fermi} detected and optically selected quasars.
We excluded 3 objects because S11 do not 
provide a reliable black hole mass estimate.
Therefore, the quasar sample under study includes 
46 objects, that are listed in Table \ref{dr7lat}.
Note that requiring that the objects have a broad emission line
with measurable FWHM automatically excludes BL Lac objects with very weak
or no emission lines.

\subsection{Plotkin et al. (2011) BL Lac sample}

We included in our analysis all the optically selected BL Lacs that are present in the work by P11.
In their work, P11 start from an original sample of 723 BL Lac candidates with EW$<$5 \AA, 
that are included in the DR7 general catalog (for selection details, see Plotkin et al.\ 2010). 
From this original sample, they selected 143 BL Lac candidates 
with reliable redshift limited to $z<0.4$, 
that match to a FIRST and/or NVSS radio source and are radio--loud
(FIRST stands for Faint Images of the Radio Sky at 20 cm, White et al. 1997;
NVSS stands for NRAO VLA Sky Survey, Condon et al. 1998).
Then they apply a spectral decomposition in order to 
separate the galaxy and AGN spectral components.
Only 71 out of the 143 BL Lacs could be successfully decomposed.
For this smaller sample of 71 BL Lacs, black hole masses 
are derived from the $M$--$\sigma_*$ relation.

A similar study on black hole masses of SDSS BL Lac objects 
has been performed in a work by Le\'on--Tavares et al.\ (2011).
The authors started from a sample of BL Lacs included in 
the SDSS Data Release 5 (DR5) and radio--detected by FIRST.
This original BL Lac sample was selected by Plotkin et al.\ (2008).
Le\'on--Tavares et al.\ performed a spectral decomposition similar 
to the one in P11 on objects in the redshift range $0.06<z<0.5$.
They obtain the black hole mass estimates for 78 BL Lacs, using 
the $M$--$\sigma_*$ relation.
The results of Le\'on--Tavares et al.\ (2011) and those of 
P11 are very similar and the two works are consistent.
We then decided to use the P11 data.

We cross--correlated the P11 sample with the clean 1LAC sample.
10 BL Lacs out of the 71 are detected by {\it Fermi},
hence we have measures of their $\gamma$--ray luminosities.
For the other 61 sources, we derived an upper limit on their $\gamma$--ray fluxes,
based on the sensitivity limit of LAT for objects with $\Gamma_{\gamma}\simeq2$.
Therefore, the upper limit in flux for these 61 BL Lacs 
is fixed at $F_{ph}=5\times10^{-9}$ ph cm$^{-2}$ s$^{-1}$.
The overall BL Lac sample under study includes 71 objects, 
that are listed in Table \ref{plotktab}.

\subsection{SDSS DR6 sample}

We selected a last group of sources from the Data Release 6 (DR6) of the SDSS, 
in order to include in our analysis possible optically intermediate objects,
that may have been excluded from the S11 and P11 catalogs.
Therefore, we cross--correlated the SDSS DR6 and the clean 1LAC and obtained 
a group of 20 additional sources, not contained in the samples mentioned above.
Three of these sources have no reliable redshifts, and we excluded them from our sample.
Since for this group of sources we do not have a black hole mass estimate,
we have chosen to assign them an average value of $M=5\times10^{8}M_\odot$.
This last group of intermediate blazars  
includes 17 objects, that are listed in Table \ref{dr6tab}.

\vskip 0.3 cm
To sum up, the sample on which we work is composed by:
\begin{itemize}

\item 46 {\it Fermi}--detected, optically selected FSRQs from S11.
These objects have detection both on $L_{\rm BLR}$ and $L_\gamma$, and 
have black hole masses estimated by S11.

\item 10 {\it Fermi}--detected, optically selected BL Lacs from P11.
Because of their original selection, these sources do not show any 
emission line. 
Therefore, we calculated the upper limit on $L_{\rm BLR}$, while  
detections are available for $L_\gamma$.
The 1LAC catalog provides for one of these sources an upper limit 
instead of a detection.
P11 provide mass estimates for all these objects.

\item 61 optically selected BL Lacs from P11 that are not
{\it Fermi}--detected.
As the other 10 sources from P11, they have mass estimates (from P11) 
and upper limits on $L_{\rm BLR}$, 
but in addition we calculated upper limits on $L_\gamma$, too.

\item 14 {\it Fermi}--detected objects that are included in the DR6 
general sample.
These objects do not show broad emission lines in their spectra, hence 
we calculated upper limits on their $L_{\rm BLR}$.
In DR6 there are no estimates of the black hole mass, hence we assigned to these objects 
an average mass value ($M=5\times10^8M_\odot$).

\item 3 {\it Fermi}--detected blazars that are included in the DR6 sample.
These objects show at least one emission line in their spectra, hence 
they have detections on both $L_{\rm BLR}$ and $L_\gamma$.
For one of them, we calculated a mass estimate from the FWHM of the H$\beta$ line, 
while we assigned to the others an average mass value.
\end{itemize}
In total, we have 49 objects with detections on both 
$L_{\rm BLR}$ and $L_\gamma$, 23 with upper limits on $L_{\rm BLR}$ and 
detections on $L_\gamma$ and 62 with upper limits on both 
luminosities.
In the following, when discussing the relation between the
BLR and the $\gamma$--ray luminosity, we will add to our sample
other 30 blazars studied in G11, 
that are listed in Table \ref{GGtab}.
Of these 30 objects, 14 are FSRQs and 16 are BL Lacs
(12 LBLs and 4 HBLs). 
29 have measured $L_{\rm BLR}$ and $\gamma$--ray detections, 
while one has an upper limit on $L_{\rm BLR}$ and a 
$\gamma$--ray detection.
The total number of blazars with measured $L_{\rm BLR}$, $L_\gamma$
and black hole mass is therefore 78. 

\begin{table*} 
\centering
\begin{tabular}{l l l l l l l l}
\hline
\hline
Name  &RA &DEC &$z$ &$\log M/M_\odot$ &Lines &$L_{\rm BLR}$ &$L_{\gamma}$ \\
      &   &    &    &                 &      &[$10^{42}$ erg/s] & $[10^{45}$ erg/s]\\
~[1]      &[2] &[3] &[4] &[5] &[6] &[7] &[8]\\
\hline   
CGRaBS J0011+0057 &00 11 30.40 & +00 57 51.7 & 1.493 & 8.95 & MgII		  & 472.91  & 217.51  \\
B3 0307+380	     &03 10 49.87 & +38 14 53.8 & 0.816 & 8.23 & H$\beta$ MgII	  & 65.97   & 54.97   \\
B2 0743+25	     &07 46 25.87 & +25 49 02.1 & 2.978 & 9.59 & CIV		  & 4199.17 & 3849.95 \\
OJ 535		     &08 24 47.24 & +55 52 42.6 & 1.418 & 9.42 & MgII		  & 2004.82 & 643.44  \\
B2 0827+24	     &08 30 52.08 & +24 10 59.8 & 0.941 & 9.01 & MgII		  & 974.10  & 227.73  \\
PKS 0906+01	     &09 09 10.09 & +01 21 35.6 & 1.025 & 9.32 & MgII		  & 1883.33 & 462.35  \\
0917+444	         &09 20 58.46 & +44 41 54.0 & 2.188 & 9.25 & MgII CIV 	  & 7075.62 & 8261.87 \\
0917+62 	         &09 21 36.23 & +62 15 52.1 & 1.453 & 9.37 & MgII		  & 978.32  & 349.93  \\
B2 0920+28	     &09 23 51.52 & +28 15 25.1 & 0.744 & 8.80 & H$\beta$ MgII	  & 257.91  & 52.06 \\
CGRaBS J0937+5008 &09 37 12.32 & +50 08 52.1 & 0.275 & 8.29 & H$\alpha$ H$\beta$ & 18.18   & 4.59 \\
CGRaBS J0941+2728 &09 41 48.11 & +27 28 38.8 & 1.306 & 8.68 & MgII		  & 4843.97 & 119.81 \\
CRATES J0946+1017 &09 46 35.06 & +10 17 06.1 & 1.005 & 8.52 & MgII		  & 639.52  & 91.52 \\
CGRaBS J0948+0022 &09 48 57.31 & +00 22 25.5 & 0.584 & 7.77 & H$\beta$ MgII	  & 126.83  & 101.32 \\
B2 0954+25A 	     &09 56 49.87 & +25 15 16.0 & 0.708 & 9.34 & H$\beta$ MgII	  & 789.15  & 32.10 \\
4C +55.17 	     &09 57 38.18 & +55 22 57.7 & 0.899 & 8.96 & MgII		  & 374.96  & 453.58 \\
CRATES J1016+0513 &10 16 03.13 & +05 13 02.3 & 1.713 & 9.11 & MgII CIV 	  & 463.36  & 2002.10 \\
B3 1030+415	     &10 33 03.70 & +41 16 06.2 & 1.116 & 8.65 & MgII		  & 857.50  & 145.07 \\
CRATES J1112+3446 &11 12 38.77 & +34 46 39.0 & 1.955 & 9.04 & MgII CIV 	  & 2108.88 & 583.93 \\
CRATES J1117+2014 &11 17 06.25 & +20 14 07.3 & 0.137 & 8.62 & $H\alpha$ H$\beta$ & 1.37    & 1.14 \\
B2 1144+40	     &11 46 58.29 & +39 58 34.2 & 1.088 & 8.98 & MgII		  & 1171.13 & 124.91 \\
4C +29.45  	     &11 59 31.83 & +29 14 43.8 & 0.724 & 9.18 & H$\beta$ MgII	  & 513.10  & 196.10 \\
CRATES J1208+5441 &12 08 54.24 & +54 41 58.1 & 1.344 & 8.67 & MgII		  & 321.88  & 333.08 \\
CRATES J1209+1810 &12 09 51.76 & +18 10 06.8 & 0.850 & 8.94 & H$\beta$ MgII	  & 288.94  & 40.69 \\
4C +04.42	     &12 22 22.55 & +04 13 15.7 & 0.965 & 8.24 & MgII		  & 720.10  & 169.22 \\
4C +21.35	     &12 24 54.46 & +21 22 46.3 & 0.433 & 8.87 & H$\beta$ MgII	  & 1617.49 & 29.89 \\
CRATES J1228+4858 &12 28 51.76 & +48 58 01.2 & 1.722 & 9.22 & MgII CIV 	  & 585.70  & 468.20 \\
CRATES J1239+0443 &12 39 32.75 & +04 43 05.3 & 1.760 & 8.67 & MgII CIV 	  & 912.84  & 1418.40 \\
B2 1255+32	     &12 57 57.23 & +32 29 29.2 & 0.805 & 8.74 & H$\beta$ MgII	  & 349.38  & 27.59 \\
B2 1308+32	     &13 10 28.66 & +32 20 43.7 & 0.997 & 8.80 & MgII		  & 837.76  & 497.26 \\
B2 1315+34A	     &13 17 36.49 & +34 25 15.8 & 1.054 & 9.29 & MgII		  & 1175.14 & 55.36 \\
CGRaBS J1321+2216 &13 21 11.20 & +22 16 12.1 & 0.948 & 8.42 & MgII		  & 272.05  & 58.82 \\
B2 1324+22	     &13 27 00.86 & +22 10 50.1 & 1.403 & 9.24 & MgII		  & 786.65  & 519.85 \\
B3 1330+476	     &13 32 45.23 & +47 22 22.6 & 0.669 & 8.56 & H$\beta$ MgII	  & 256.41  & 18.89 \\
B2 1348+30B	     &13 50 52.73 & +30 34 53.5 & 0.712 & 8.69 & H$\beta$ MgII	  & 211.68  & 22.73 \\
PKS 1434+235	     &14 36 40.98 & +23 21 03.2 & 1.547 & 8.44 & MgII CIV 	  & 595.91  & 110.86 \\
PKS 1502+106	     &15 04 24.98 & +10 29 39.1 & 1.839 & 9.64 & MgII CIV 	  & 1983.07 & 22563.8 \\
PKS 1509+022	     &15 12 15.74 & +02 03 16.9 & 0.219 & 8.84 & H$\alpha$ H$\beta$ & 10.56   & 3.98 \\
PKS 1546+027	     &15 49 29.43 & +02 37 01.1 & 0.414 & 8.61 & H$\beta$ MgII	  & 821.22  & 22.16 \\
4C +05.64	     &15 50 35.27 & +05 27 10.4 & 1.417 & 9.38 & MgII		  & 1138.94 & 209.33 \\
PKS 1551+130	     &15 53 32.69 & +12 56 51.7 & 1.308 & 9.10 & MgII		  & 1587.17 & 1003.15 \\
4C +10.45	     &16 08 46.20 & +10 29 07.7 & 1.231 & 8.64 & MgII		  & 1014.70 & 361.88 \\
B2 1611+34	     &16 13 41.06 & +34 12 47.8 & 1.399 & 9.12 & MgII		  & 3131.09 & 95.51 \\
CRATES J1616+4632 &16 16 03.77 & +46 32 25.2 & 0.950 & 8.44 & MgII		  & 233.23  & 93.91 \\
4C +38.41 	     &16 35 15.49 & +38 08 04.4 & 1.813 & 9.53 & MgII CIV 	  & 5743.01 & 3420.04 \\
CRATES J2118+0013 &21 18 17.39 & +00 13 16.7 & 0.462 & 7.93 & H$\beta$ MgII	  & 114.78  & 6.23 \\
PKS 2227--08	     &22 29 40.08 & --08 32 54.4 & 1.559 & 8.95 & MgII CIV 	  & 4613.63 & 2464.28 \\
\hline
\hline 
\end{tabular}
\vskip 0.4 true cm
\caption{Sources from the DR7 Quasar Catalog that are present in the 1LAC {\it Fermi} sample.
Col. [1]: name;
Col. [2]: right ascension;
Col. [3]: declination;
Col. [4]: redshift;
Col. [5]: Logarithm of the black hole mass (in solar masses, best estimate from S11);
Col. [6]: lines measured by S11, from which $L_{\rm BLR}$ has been derived;
Col. [7]: Broad Line Region luminosity ($10^{42}$ erg s$^{-1}$), obtained  from 
         the line luminosities calculated by S11; 
Col. [8]: $\gamma$--ray luminosity from {\it Fermi} data ($10^{45}$ erg s$^{-1}$), averaged on the 
         first 11 months of {\it Fermi} operations.
}
\label{dr7lat}
\end{table*}

\begin{table*} 
\centering
\begin{tabular}{l l l l l l l }
\hline
\hline
Name (SDSS J\ldots)  &$z$ &$\log M/M_\odot$ &Lines &UL $L_{\rm BLR}$ &UL $L_{\gamma}$ &$L_{\gamma}$ \\
                     &    &                 &	   &[$10^{42}$ erg s$^{-1}$] 
                     &[$10^{44}$ erg s$^{-1}$] &[$10^{44}$ erg s$^{-1}$] \\
~[1]      &[2] &[3] &[4] &[5] &[6] &[7] \\
\hline   
002200.95 +000657.9 & 0.306 & 8.49 &H$\alpha$ H$\beta$   & 3.65  & 1.07  &	  \\
005620.07 --093629.7& 0.103 & 9.01 &H$\alpha$ H$\beta$   & 1.02  & 0.95  &	  \\
075437.07 +391047.7 & 0.096 & 8.24 &H$\alpha$ H$\beta$   & 0.52  & 0.81  &	  \\
080018.79 +164557.1 & 0.309 & 8.58 &H$\alpha$ H$\beta$   & 5.68  & 1.09  &	  \\
082323.24 +152447.9 & 0.167 & 8.80 &H$\alpha$ H$\beta$   & 1.62  & 2.69  &	  \\
082814.20 +415351.9 & 0.226 & 8.83 &H$\alpha$ H$\beta$   & 2.55  & 5.24  &	  \\
083417.58 +182501.6 & 0.336 & 9.34 &H$\beta$    	 & 5.79  & 13.18 &	 \\
083548.14 +151717.0 & 0.168 & 7.94 &H$\alpha$ H$\beta$   & 2.87  & 2.75  &	  \\
083918.74 +361856.1 & 0.335 & 8.50 &H$\beta$    	 & 7.89  & 13.18 &	 \\
084712.93 +113350.2 & 0.198 & 8.52 &H$\alpha$ H$\beta$   & 3.38  &	 & 3.34   \\
085036.20 +345522.6 & 0.145 & 8.61 &H$\alpha$ H$\beta$   & 2.07  &	 & 1.74   \\
085729.78 +062725.0 & 0.338 & 8.23 &H$\beta$    	 & 7.23  & 13.48 &	 \\
085749.80 +013530.3 & 0.281 & 8.69 &H$\alpha$ H$\beta$   & 5.85  & 8.70  &	  \\
090207.95 +454433.0 & 0.289 & 8.78 &H$\alpha$ H$\beta$   & 5.24  & 9.33  &	  \\
090314.70 +405559.8 & 0.188 & 8.28 &H$\alpha$ H$\beta$   & 2.02  & 3.54  &	  \\
090953.28 +310603.1 & 0.272 & 8.95 &H$\alpha$ H$\beta$   & 6.54  & 8.12  &	  \\
091045.30 +254812.8 & 0.384 & 8.51 &H$\beta$    	 & 9.92  & 18.19 &	 \\
091651.94 +523828.3 & 0.190 & 8.53 &H$\alpha$ H$\beta$   & 3.07  & 3.63  &	  \\
093037.57 +495025.6 & 0.187 & 8.48 &H$\alpha$ H$\beta$   & 3.26  & 3.46  &	  \\
094022.44 +614826.1 & 0.211 & 8.57 &H$\alpha$ H$\beta$   & 3.55  &	 & 11.57  \\
094542.23 +575747.7 & 0.229 & 8.63 &H$\alpha$ H$\beta$   & 3.23  &	 & 15.97  \\
101244.30 +422957.0 & 0.365 & 8.67 &H$\beta$    	 & 13.30 & 15.84 &	 \\
102453.63 +233234.0 & 0.165 & 7.46 &H$\alpha$ H$\beta$   & 2.68  & 2.63  &	  \\
102523.04 +040228.9 & 0.208 & 8.18 &H$\alpha$ H$\beta$   & 2.74  & 4.36  &	  \\
103317.94 +422236.3 & 0.211 & 8.59 &H$\alpha$ H$\beta$   & 3.18  & 4.57  &	  \\
104029.01 +094754.2 & 0.304 & 8.70 &H$\alpha$ H$\beta$   & 6.45  & 10.47 &	  \\
104149.15 +390119.5 & 0.208 & 8.55 &H$\alpha$ H$\beta$   & 3.23  & 4.36  &	  \\
104255.44 +151314.9 & 0.307 & 7.81 &H$\alpha$ H$\beta$   & 4.56  & 10.71 &	  \\
105344.12 +492955.9 & 0.140 & 8.47 &H$\alpha$ H$\beta$   & 1.68  &	 & 3.03   \\
105538.62 +305251.0 & 0.243 & 8.43 &H$\alpha$ H$\beta$   & 4.10  & 6.30  &	  \\
105606.61 +025213.4 & 0.236 & 8.11 &H$\alpha$ H$\beta$   & 3.43  & 5.88  &	  \\
105723.09 +230318.7 & 0.378 & 8.32 &H$\beta$    	 & 9.91  & 17.37 &	  \\
112059.74 +014456.9 & 0.368 & 9.60 &H$\beta$    	 & 9.81  & 16.21 &	  \\
113630.09 +673704.3 & 0.134 & 8.30 &H$\alpha$ H$\beta$   & 1.07  &	 & 2.03   \\
114023.48 +152809.7 & 0.244 & 9.46 &H$\alpha$ H$\beta$   & 4.43  & 6.30  &	  \\
114535.10 --034001.4& 0.168 & 8.27 &H$\alpha$ H$\beta$   & 2.08  & 2.75  &	  \\
115404.55 --001009.8& 0.254 & 8.36 &H$\alpha$ H$\beta$   & 3.54  &	 & 5.53   \\
115709.53 +282200.7 & 0.300 & 9.20 &H$\alpha$ H$\beta$   & 5.60  & 10.23 &	  \\
 \ldots & & & & & & \\
\hline 
\end{tabular}
\vskip 0.4 true cm
\caption{BL Lacs from the work by P11.
Col. [1]: SDSS name;
Col. [2]: redshift;
Col. [3]: Logarithm of the black hole mass (P11, in solar masses);
Col. [4]: lines from which we derived the upper limits, as described in Section \ref{UL};
Col. [5]: upper limit on the Broad Line Region luminosity ($10^{42}$ erg s$^{-1}$), 
	  obtained from the UL on line fluxes;
Col. [6]: upper limit on the $\gamma$--ray luminosity obtained from the \textit{Fermi}--LAT sensitivity limit,
          in units of $10^{44}$ erg s$^{-1}$;
Col. [7]: $\gamma$--ray luminosity from {\it Fermi} data ($10^{44}$ erg s$^{-1}$), averaged on the 
	  first 11 months of {\it Fermi} operations.
}
\label{plotktab}
\end{table*}

\begin{table*} 
\setcounter{table}{1}
\centering
\begin{tabular}{l l l l l l l }
\hline
\hline
Name (SDSS J\ldots)  &$z$ &$\log M/M_\odot$ &Lines &UL $L_{\rm BLR}$ &UL $L_{\gamma}$ &$L_{\gamma}$ \\
                     &    &                 &	   &[$10^{42}$ erg s$^{-1}$] 
                     &[$10^{44}$ erg s$^{-1}$] &[$10^{44}$ erg s$^{-1}$] \\
~[1]      &[2] &[3] &[4] &[5] &[6] &[7] \\
\hline 
   \ldots & & & & & & \\
120837.27 +115937.9 & 0.369 & 8.66 &H$\beta$	       & 12.56 & 16.21 &	\\
123123.90 +142124.4 & 0.256 & 8.62 &H$\alpha$ H$\beta$ & 5.37 & 7.07  &        \\
123131.39 +641418.2 & 0.163 & 8.84 &H$\alpha$ H$\beta$ & 1.96 & 2.57  &        \\
123831.24 +540651.8 & 0.224 & 8.61 &H$\alpha$ H$\beta$ & 4.15 & 5.24  &        \\
125300.95 +382625.7 & 0.371 & 8.24 &H$\beta$	       & 6.81 & 16.59 &        \\
131330.12 +020105.9 & 0.356 & 8.50 &H$\beta$	       & 8.05 & 15.13 &        \\
132231.46 +134429.8 & 0.377 & 8.97 &H$\beta$	       & 9.92 & 17.37 &        \\
132239.31 +494336.2 & 0.332 & 8.67 &H$\beta$	       & 7.66 & 12.88 &        \\
132301.00 +043951.3 & 0.224 & 8.86 &H$\alpha$ H$\beta$ & 4.23 & 5.24  &        \\
132617.70 +122958.7 & 0.204 & 8.63 &H$\alpha$ H$\beta$ & 3.66 & 4.16  &        \\
133612.16 +231958.0 & 0.267 & 8.56 &H$\alpha$ H$\beta$ & 4.70 & 7.76  &        \\
134105.10 +395945.4 & 0.172 & 8.48 &H$\alpha$ H$\beta$ & 2.79 & 10.12$^*$ &    \\
134136.23 +551437.9 & 0.207 & 8.29 &H$\alpha$ H$\beta$ & 3.68 & 4.36  &        \\
134633.98 +244058.4 & 0.167 & 8.29 &H$\alpha$ H$\beta$ & 2.42 & 2.69  &        \\
135314.08 +374113.9 & 0.216 & 8.79 &H$\alpha$ H$\beta$ & 3.66 & 4.78  &        \\
140350.28 +243304.8 & 0.343 & 8.39 &H$\beta$	       & 7.44 & 13.80 &        \\
142421.17 +370552.8 & 0.290 & 8.39 &H$\alpha$ H$\beta$ & 5.34 & 9.33  &        \\
142832.60 +424021.0 & 0.129 & 8.70 &H$\alpha$ H$\beta$ & 1.90 &       & 1.86   \\
144248.28 +120040.2 & 0.163 & 8.94 &H$\alpha$ H$\beta$ & 2.46 &       & 3.45   \\
144932.70 +274621.6 & 0.227 & 8.86 &H$\alpha$ H$\beta$ & 3.77 & 5.37  &        \\
153311.25 +185429.1 & 0.307 & 8.91 &H$\alpha$ H$\beta$ & 6.45 & 10.71 &        \\
155412.07 +241426.6 & 0.301 & 8.59 &H$\alpha$ H$\beta$ & 5.00 & 10.23 &        \\
155424.12 +201125.4 & 0.222 & 8.94 &H$\alpha$ H$\beta$ & 3.66 & 5.12  &        \\
160118.96 +063136.0 & 0.358 & 8.69 &H$\beta$	       & 6.78 & 15.13 &        \\
160519.04 +542059.9 & 0.212 & 7.85 &H$\alpha$ H$\beta$ & 3.28 & 4.57  &        \\
161541.21 +471111.7 & 0.199 & 8.17 &H$\alpha$ H$\beta$ & 3.92 & 3.98  &        \\
161706.32 +410647.0 & 0.267 & 7.84 &H$\alpha$ H$\beta$ & 6.22 & 7.76  &        \\
162839.03 +252755.9 & 0.220 & 8.90 &H$\alpha$ H$\beta$ & 2.71 & 5.01  &        \\
163726.66 +454749.0 & 0.192 & 8.42 &H$\alpha$ H$\beta$ & 3.07 & 3.71  &        \\
164419.97 +454644.3 & 0.225 & 8.76 &H$\alpha$ H$\beta$ & 4.64 & 5.24  &        \\
205456.85 +001537.7 & 0.151 & 8.67 &H$\alpha$ H$\beta$ & 1.49 & 2.18  &        \\
205938.57 --003756.0 & 0.335 &7.16 &H$\beta$	       & 6.61 & 13.18 &        \\
223301.11 +133602.0 & 0.214 & 8.54 &H$\alpha$ H$\beta$ & 3.02 & 4.67  &        \\
\hline
\hline 
\end{tabular}
\vskip 0.4 true cm
\caption{BL Lacs from the work by P11. $^*$: the upper limit is from the
Abdo et al. (2010a) list.
Col. [1]: SDSS name;
Col. [2]: redshift;
Col. [3]: Logarithm of the black hole mass (P11, in solar masses);
Col. [4]: lines from which we derived the upper limits, as described in Section \ref{UL};
Col. [5]: upper limit on the Broad Line Region luminosity ($10^{42}$ erg s$^{-1}$), 
	  obtained from the UL on line fluxes;
Col. [6]: upper limit on the $\gamma$--ray luminosity obtained from the \textit{Fermi}--LAT sensitivity limit,
          in units of $10^{44}$ erg s$^{-1}$;
Col. [7]: $\gamma$--ray luminosity from {\it Fermi} data ($10^{44}$ erg s$^{-1}$), averaged on the 
	  first 11 months of {\it Fermi} operations.
}
\label{plotktab}
\end{table*}

\begin{table*} 
\centering
\begin{tabular}{l l l l l l l l}
\hline
\hline
Name  &RA &DEC &$z$ &Lines &UL $L_{\rm BLR}$ &$L_{\rm BLR}$ &$L_{\gamma}$ \\
      &    &    &   &       &[$10^{42}$ erg s$^{-1}$] &[$10^{42}$ erg s$^{-1}$] &[$10^{45}$ erg s$^{-1}$] \\
~[1]  &[2] &[3] &[4] &[5]  &[6] 		&[7] 		&[8] \\
\hline
B2 0806+35          &08 09 38.87 & +34 55 37.2 & 0.082 &H$\alpha$ H$\beta$ &0.49   &	  & 0.18   \\  
CRATES J0809+5218   &08 09 49.18 & +52 18 58.2 & 0.138 &H$\alpha$ H$\beta$ &7.581  &	  & 1.45   \\  
Ton 1015            &09 10 37.03 & +33 29 24.4 & 0.354 &H$\alpha$ H$\beta$ &30.675 &	  & 4.54   \\  
CRATES J1012+0630   &10 12 13.34 & +06 30 57.2 & 0.727 &H$\beta$ MgII	   &134.295&	  & 35.9  \\   
1ES 1011+496        &10 15 04.14 & +49 26 00.6 & 0.212 &H$\beta$	   &19.9   &	  & 9.72   \\  
B2 1040+24A         &10 43 09.04 & +24 08 35.4 & 0.560 &MgII		   &	   &25.4  & 13.7   \\  
PKS 1055+01         &10 58 29.60 & +01 33 58.8 & 0.890 &MgII		   &	   &131.8 & 377.2 \\   
CGRaBS J1058+5628   &10 58 37.73 & +56 28 11.1 & 0.143 &H$\alpha$ H$\beta$ &3.85   &	  & 3.58   \\  
PKS 1106+023$^{a}$  &11 08 45.48 & +02 02 40.8 & 0.157 &H$\beta$	   &	   &4.7   & 0.51    \\ 
1ES 1118+424        &11 20 48.06 & +42 12 12.4 & 0.124 &H$\alpha$ H$\beta$ &1.797  &	  & 0.47   \\  
B2 1147+24          &11 50 19.21 & +24 17 53.8 & 0.200 &H$\alpha$ H$\beta$ &9.627  &	  & 1.47   \\  
B2 1218+30          &12 21 21.94 & +30 10 37.2 & 0.184 &H$\alpha$ H$\beta$ &6.04   &	  & 3.34   \\  
W Com               &12 21 31.69 & +28 13 58.4 & 0.102 &H$\alpha$ H$\beta$ &5.26   &	  & 1.80   \\  
B2 1229+29          &12 31 43.57 & +28 47 49.7 & 0.236 &H$\alpha$ H$\beta$ &9.696  &	  & 4.31   \\  
CRATES J1253+0326   &12 53 47.00 & +03 26 30.3 & 0.066 &H$\alpha$ H$\beta$ &0.41   &	  & 0.13   \\  
PG 1437+398         &14 39 17.47 & +39 32 42.8 & 0.344 &H$\alpha$ H$\beta$ &17.93  &	  & 4.27   \\  
\hline
\hline 
\end{tabular}
\vskip 0.4 true cm
\caption{Sources from the DR6 Catalog that are present in the 1LAC Fermi sample.
Col. [1]: name;
Col. [2]: right ascension;
Col. [3]: declination;
Col. [4]: redshift;
Col. [5]: lines measured or for which we derived the upper limits as described in Section \ref{UL};
Col. [6]: upper limit on the Broad Line Region luminosity ($10^{42}$ erg s$^{-1}$), 
obtained from the UL on line fluxes; 
Col. [7]: luminosity of the Broad Line region ($10^{42}$ erg s$^{-1}$);
Col. [8]: $\gamma$--ray luminosity from {\it Fermi} data ($10^{45}$ erg s$^{-1}$), averaged on the 
first 11 months of {\it Fermi} operations.
The black masses are not available, hence a medium mass value 
($M=5\times 10^{8} M_\odot$) has been assigned to all of them.
$^{a}$: PKS 1106+023 has the H$\beta$ line measured in DR6 Catalog, hence its mass can be 
estimated by the Chiaberge \& Marconi relation (2011):
a value of $M_=4\times 10^{7} M_\odot$ is obtained.
}
\label{dr6tab}
\end{table*}

\begin{table*} 
\centering
\begin{tabular}{l l l l l l l }
\hline
\hline
Name 		&RA  &DEC    &$z$   &$\log M/M_\odot$ &$L_{\rm BLR}$ &$L_{\gamma}$ \\
             &   &       &      &                  &[$10^{42}$ erg s$^{-1}$] &[$10^{45}$ erg s$^{-1}$] \\
~[1]      &[2] &[3] &[4] &[5] &[6] &[7] \\
\hline   
{\bf `FS'}	&	     &	          &      &    &	    &	    \\
PKS 0208-512	&02 11 13.18 &+10 51 34.8 &1.003 &9.2 &3700 &489.8  \\
PKS 0235+164	&02 38 38.93 &+16 36 59.3 &0.940 &9.0 &100  &1737.8 \\
PKS 0426-380	&04 28 40.42 &-37 56 19.6 &1.111 &8.6 &110  &1513.6 \\
PKS 0537-441	&05 38 50.35 &-44 05 08.7 &0.892 &8.8 &690  &1000.0 \\
PKS 0808+019	&08 11 26.71 &+01 46 52.2 &1.148 &8.5 &42   &120.2  \\
\hline
{\bf LBL}	&	     &	          &      &    &     &	    \\
PKS 0521-36	&05 22 57.98 &-36 27 30.9 &0.055 &8.6 &4.8  &0.28   \\
PKS 0829+046	&08 31 48.88 &+04 29 39.1 &0.174 &8.8 &3.7  &2.45	 \\
OJ 287 	        &08 54 48.87 &+20 06 30.6 &0.306 &8.8 &6.8  &1.51    \\  
TXS 0954+658	&09 58 47.25 &+65 33 54.8 &0.367 &8.5 &2.8  &4.90	 \\
PMN 1012+0630	&10 12 13.35 &+06 30 57.2 &0.727 &8.5 &7.8  &35.5	 \\
PKS 1057-79	&10 58 43.40 &-80 03 54.2 &0.581 &8.8 &58   &45.7	 \\
PKS 1519-273	&15 22 37.68 &-27 30 10.8 &1.294 &8.8 &34   &354.8  \\
PKS 1749+096	&17 51 32.82 &+09 39 00.7 &0.322 &8.7 &50   &26.9	 \\
S5 1803+78	&18 00 45.68 &+78 28 04.0 &0.680 &8.6 &710  &87.1	 \\
3C 371		&18 06 50.68 &+69 49 28.1 &0.050 &8.7 &1.0  &0.19	 \\
BL Lac		&22 02 43.29 &+42 16 40.0 &0.069 &8.7 &3.3  &0.93	 \\
PKS 2240-260    &22 43 26.47 &-25 44 31.4 &0.774 &8.6 &29   &56.23   \\
\hline
{\bf HBL}	&	     &	          &      &    &     &	 \\
Mkn 421 	&11 04 27.30 &+38 12 32.0 &0.031 &8.5 &0.5  &0.33	 \\
Mkn 501 	&16 53 52.20 &+39 45 37.0 &0.034 &9.0 &1.6  &0.09	 \\
PKS 2005-489	&20 09 25.40 &-48 49 54.0 &0.071 &8.5 &1.1  &0.32	 \\
WGA 1204.2-0710 &12 04 16.66 &-07 10 09.0 &0.185 &8.8 &$<$9.5 &0.98	 \\
\hline
{\bf FSRQ}	&	     &	          &      &    &     &	 \\
TXS 1013+054	&10 16 03.10 &+05 13 02.0 &1.713 &9.5 &889  &1584.9 \\
S4 1030+61	&10 33 51.40 &+60 51 07.3 &1.401 &9.5 &450  &741.31  \\
PKS 1144-379	&11 47 01.40 &-38 12 11.0 &1.049 &8.5 &400  &223.9  \\
3C 273		&12 29 06.69 &+02 03 08.5 &0.158 &8.9 &3380 &21.4	 \\
3C 279		&12 56 11.10 &-05 47 21.5 &0.536 &8.9 &242  &204.2  \\
PKS 1510-089	&15 12 50.50 &-09 06 00.0 &0.360 &8.6 &741  &125.9  \\
OX 169		&21 43 35.50 &+17 43 48.6 &0.213 &8.6 &182  &8.51	 \\
CTA102		&22 32 36.40 &+11 43 53.8 &1.037 &8.7 &4140 &489.8  \\
3C 454.3	&22 53 57.70 &+16 08 53.6 &0.859 &8.7 &3330 &5011.9 \\
\hline
\hline 
\end{tabular}
\vskip 0.4 true cm
\caption{Blazars from G11.
Col. [1]: name;
Col. [2]: right ascension;
Col. [3]: declination;
Col. [4]: redshift;
Col. [5]: Logarithm of the black hole mass (in solar masses);
Col. [6]: Broad Line Region luminosity ($10^{42}$ erg s$^{-1}$);
Col. [7]: $\gamma$--ray luminosity ($10^{45}$ erg s$^{-1}$).
}
\label{GGtab}
\end{table*}


%
\section{Broad Line Luminosities}

We have taken the luminosity of the emission lines of the 
blazars in the SDSS DR7 Quasar sample directly from the listed values in the
S11 catalog. 
For calculating the total luminosity of the broad lines, we have followed
Celotti, Padovani \& Ghisellini (1997). 
Specifically we set the Ly$\alpha$ 
flux contribution to 100, the relative weight of H$\alpha$, H$\beta$, MgII and 
CIV lines respectively to 77, 22, 34 and 63 (see Francis et al. 1991).
The total broad line flux is fixed at $555.76$.
The $L_{\rm BLR}$ value or upper limit of each source has been derived
using these proportions.
In Tab. \ref{dr7lat} we list these blazars reporting the type of lines used
for calculating $L_{\rm BLR}$ and the values of the estimated $L_{\rm BLR}$ and
the observed $L_\gamma$.
When more than one line is present, we calculate the simple average of
the $L_{\rm BLR}$
estimated from each line.

\subsection{Upper limits on  the broad line luminosity}
\label{UL}
While the SDSS DR7 Quasar sample is selected in order 
to contain spectra with prominent broad emission lines, that have
been measured by S11, the other two samples include mostly lineless objects.
In these cases, we need to derive upper limits on the line fluxes 
(UL$_{F_{\rm line}}$).

To this aim, the observed spectrum has been fitted with a power--law
model, with the addition of possible absorption
or narrow emission
features modeled as Gaussian profiles.
Absorption lines are well visible in the spectra included in 
P11 BL Lac sample and the P11 work itself provides the variance of 
each absorption feature ($\sigma_*$).
Such a modeling includes the lineless power--law and the  
narrow features.
The broad emission line for which we want to obtain the upper limit 
is accounted as an additional Gaussian profile,
with a variable flux value $F_{\rm line}$ and a FWHM fixed at the average 
value $v_{\rm FWHM}=4000$ km s$^{-1}$,
as suggested in Decarli et al.\ (2011).
This value is an average for all blazars, and it is consistent 
with the median FWHM values that can be obtained from the whole SDSS  
Quasar sample.
Even though in the case of BL Lacs the average value is possibly slightly smaller, 
we prefer to maintain a larger average FWHM value, in order to derive 
more conservative (i.e. less stringent) upper limits. 
To define the UL$_{F_{\rm line}}$ we perform a $\chi^2$ test, 
varying the $F_{\rm line}$ value until our model returns an unacceptable fit.
Then we define the upper limit on the line flux as the $F_{\rm line}$ 
for which we obtain  $\chi^2 > \chi^2(99\%)$.
Over this critical value, the model is no more acceptable to fit the data, 
and we should actually see a broad line emerging over the continuum, 
if present.
In order to derive meaningful upper limits, 
we required a signal--to--noise ratio $S/N>5$ in the wavelenght interval 
in which we performed our analysis.
Hence we checked the 
signal--to--noise ratios of the spectra in our sample, and 
we excluded B3 1432+422 (SDSS J143405.69+420316.0, from the DR6 sample), 
because its signal--to--noise ratio was $S/N<5$ 
over the whole spectrum.
Therefore, we are left with 16 DR6 objects, listed in Table \ref{dr6tab}.

In principle, the process used to derive the upper limits could be applied to the four
lines measured in the work by S11 (i.e.\ H$\alpha$, H$\beta$, 
MgII and CIV). 
The objects included in the P11 sample all have $z<0.4$, hence
the UL$_{\rm F_{line}}$ can be derived only for the H$\alpha$ and
H$\beta$ lines.
We then derive the upper limits for these two lines.
We applied the procedure also to the low redshift objects included in the DR6 sample.
In 3 objects from this sample the redshift is sufficiently large to derive the 
upper limit on the MgII line.

\begin{figure}
\vskip -0.6 cm 
\hskip -0.4 cm
\psfig{figure=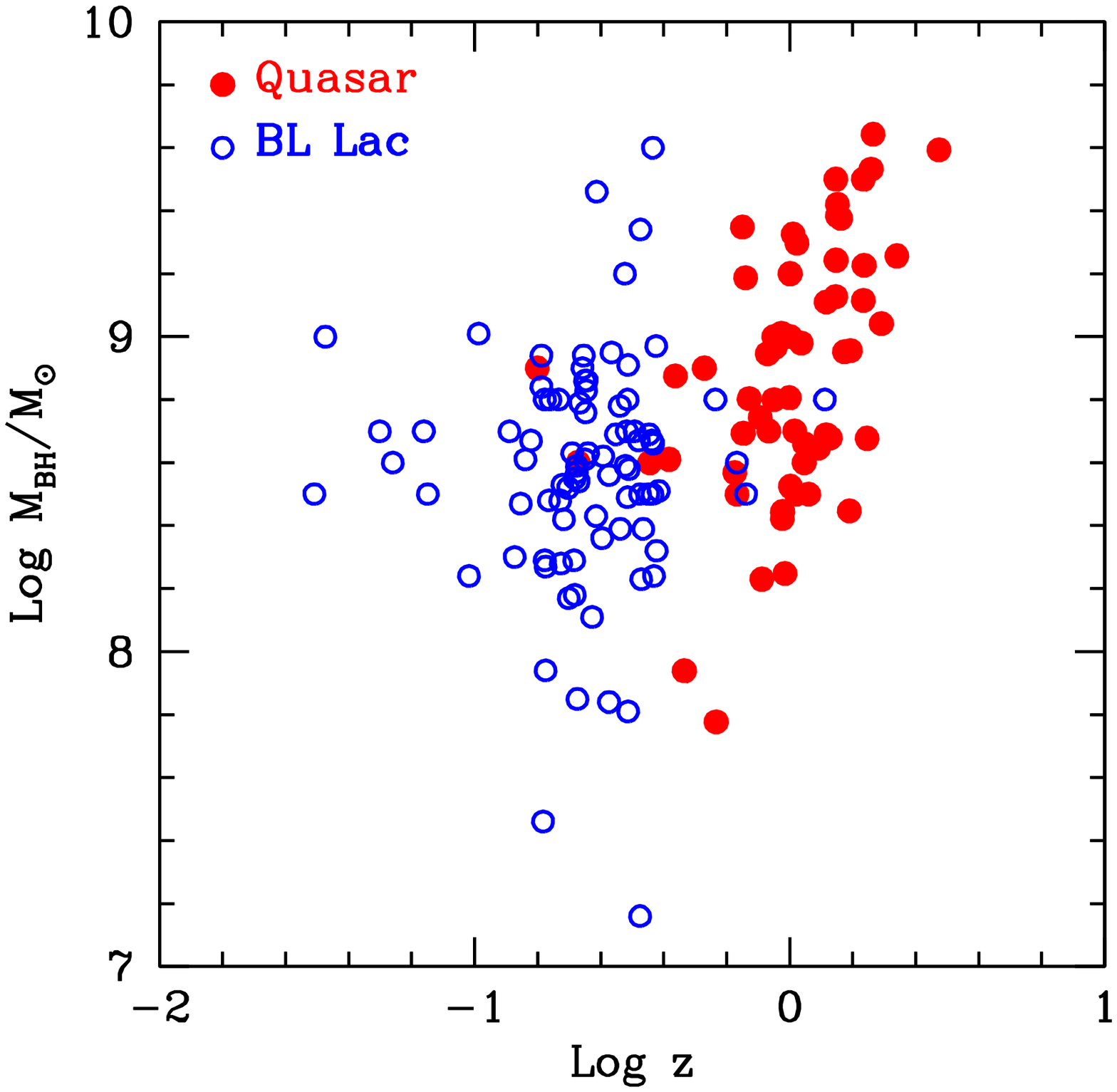,width=9cm,height=8cm}
\vskip -0.5 cm
\caption{
Black hole masses as a function of redshift. 
Empty circles are BL Lacs, filled circles are FSRQs.
BL Lacs of our sample have significantly smaller redshift than FSRQs.
}
\label{mz}
\end{figure}

\begin{figure*}
\vskip -0.6 cm 
\hskip -1. cm
\psfig{figure=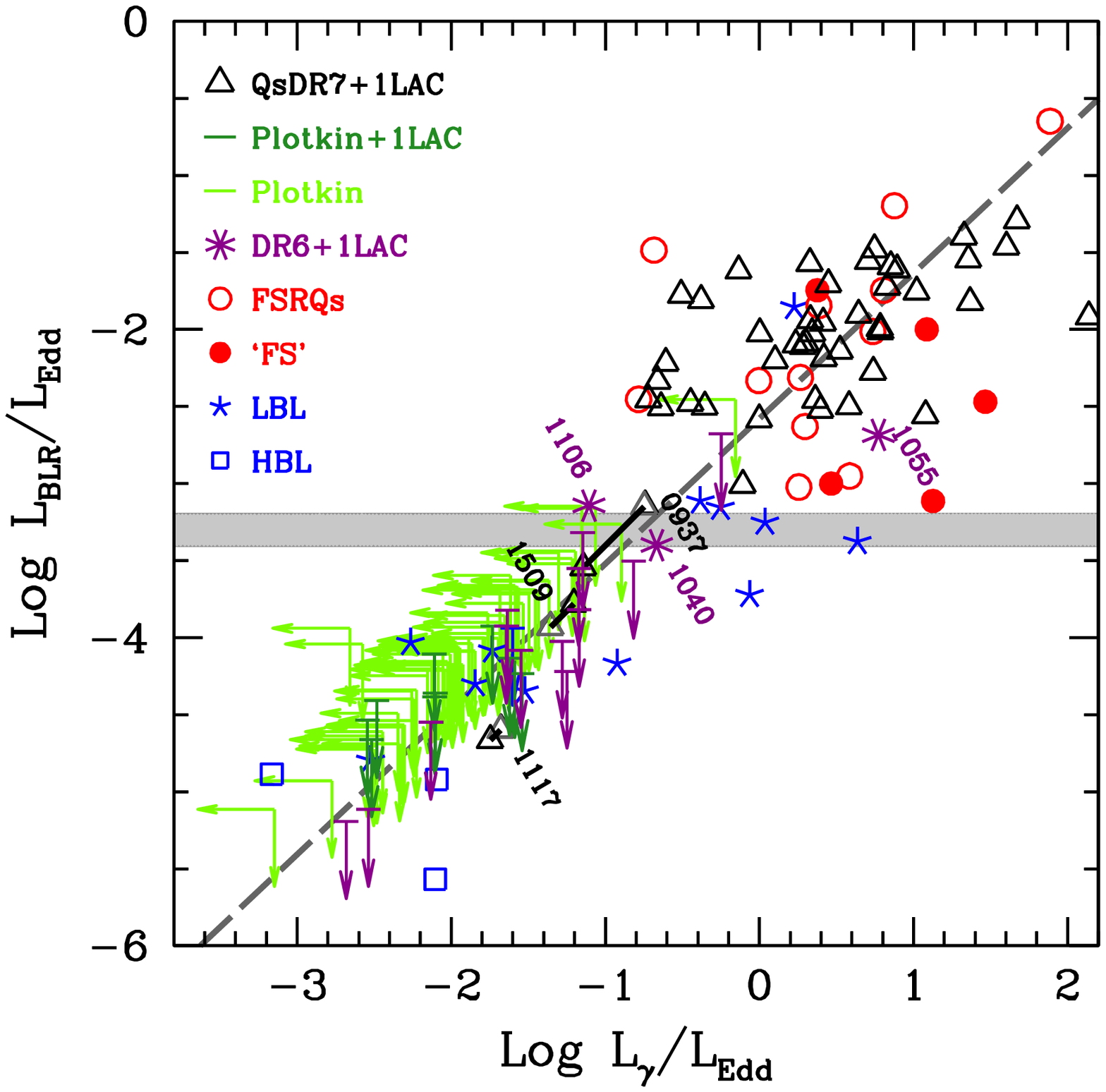,width=19cm,height=19cm}
\vskip -1 cm
\caption{Luminosity of the broad line region (in Eddington units) 
for the sources from our samples and from G11
as a function of the $\gamma$--ray luminosity (in Eddington units).
Different symbols correspond to different samples or a 
different classification of the sources, as labelled.
The three (violet) asterisks are the only sources with visible 
broad emission lines from the DR6 {\it Fermi} detected sample.
The three labelled triangles have their synchrotron emission
dominating over the thermal emission in their SEDs.
Hence, to avoid black hole mass estimate errors, possibly occurred in the
S11 automatic calculation, we also assigned them an average $M_{\rm BH}$ value
($M_{\rm BH}=5\times10^8M_\odot$).
These changes are highlighted by the thick (black) segments ending
to the black triangles (corresponding to the average $M_{\rm BH}$ value).
The grey stripe indicates the luminosity ``divide" between 
FSRQs and BL Lacs at  
$L_{\rm BLR}/L_{\rm Edd}\sim5\times10^{-4}$.
}
\label{class}
\end{figure*}

\subsection{The distribution of black hole masses}
                                
As a byproduct of our study, we have collected (from the S11 and P11 samples) a large
number of estimates for the black hole masses both in FSRQs and BL Lacs.
FSRQs show a distribution skewed towards larger masses than BL Lacs:
$\langle \log M_{\rm FSRQ}\rangle =8.88\pm 0.40$; $\langle\log M_{\rm BL \, Lac}\rangle=8.57\pm 0.37$, 
while the average of all masses is $\langle \log M_{\rm all}\rangle=8.70\pm 0.41$.
We believe that at least in part this is due to a selection effect, since most BL Lacs come
from the P11 sample, therefore they have been selected to be at $z<0.4$.
Assuming that very large black hole masses are rarer than smaller ones,
one expects that the largest masses are found only when considering 
relatively large redshifts.
This is illustrated by Fig. \ref{mz}, that shows the black hole masses as a 
function of redshift for BL Lacs and FSRQs.
Apart from few exceptions, the BL Lacs extend out to $z\sim 0.4$ 
(by construction, given the redshift limit of the P11 sample),
while FSRQs cluster around $z\sim 1$.


\begin{table*} 
\centering
\begin{tabular}{l l l l l l l }
\hline
\hline
Kendall 	&\multicolumn{2}{c}{\large $\tau$} \\
		&detections &det.+UL \\
\hline   
$\log  L_{\rm BLR}$--$\log L_\gamma$            			&0.530 &0.561  \\
$\log  L_{\rm BLR}$--$\log L_\gamma$, $z$       			&0.281 &0.386  \\
$\log (L_{\rm BLR}/L_{\rm Edd})$--$\log (L_\gamma/L_{\rm Edd})$		&0.398 &0.529  \\ 
$\log (L_{\rm BLR}/L_{\rm Edd})$--$\log (L_\gamma/L_{\rm Edd})$, $z$	&0.266 &0.376  \\ 
\hline
\hline
Minimum square fit (only det.)    &$m$ &$q$ &$N$ &$r$ &$P$ \\
\hline   
$\log  L_{\rm BLR}$--$\log L_\gamma$            &0.93 &0.61 &78 &0.83 &$<4\times 10^{-8}$  \\
$\log  L_{\rm BLR}$--$\log L_\gamma$, $z$       &0.93 &0.61 &78 &0.64 &$<4\times 10^{-8}$  \\
$\log (L_{\rm BLR}/L_{\rm Edd})$--$\log (L_\gamma/L_{\rm Edd})$           &0.94 &--2.58 &78 &0.78 &$<4\times 10^{-8}$  \\ 
$\log (L_{\rm BLR}/L_{\rm Edd})$--$\log (L_\gamma/L_{\rm Edd})$, $z$      &0.94 &--2.58 &78 &0.67 &$<4\times 10^{-8}$  \\ 
$\log (L_{\rm BLR}/L_{\rm Edd})$--$\log (L_\gamma/L_{\rm Edd})$, $z$, $M$ &0.94 &--2.58 &78 &0.64 &$<4\times 10^{-8}$  \\ 
\hline
\hline 
\end{tabular}
\vskip 0.4 true cm
\caption{
The first part of the table reports the
results of the non--parametric Kendall test for 
the complete sample, taking into consideration at first only the sources with 
both $L_{\rm BLR}$ and $L_{\gamma}$ detected, then including upper limits.
We list also the results when accounting for the common dependence on redshift.
The bottom part of the table reports the results of a partial correlation analysis using a least square fit.
We have excluded upper limits from the analysis.
Correlations are of the form $\log y = m\log x +q$. 
The listed slopes $m$ refer to the bisector (of the two correlations $x$ vs $y$ and $y$ vs $x$).
$N$ is the total number of objects. $r$ is the correlation coefficient obtained from the analysis.
$P$ is the probability that the correlation is random.
}
\label{correlation}
\end{table*}

\section{The $L_{\rm BLR}$--$L_\gamma$ relation}

Fig. \ref{class} is the key result of our work.
It shows the luminosity of the BLR as a function of the observed
$\gamma$--ray luminosity, both measured in Eddington units.
Arrows correspond to upper limits.
Different symbols correspond to blazars belonging to different samples, as labelled.
Note that we have also added the blazars studied in G11, but omitting the objects
in common.
Fig. \ref{class} shows a clear trend.
Since the range in black hole masses is relatively narrow, we obtain a similar
trend when plotting $L_{\rm BLR}$ vs $L_\gamma$.
We have quantified it first by using the Kendall non--parametric
test considering the detected sources (i.e. excluding upper limits).
The Kendall $\tau$ in this case is listed in Tab. \ref{correlation}.
The correlation is significant both considering $\log L_{\rm BLR}$ vs $\log L_{\rm \gamma}$
and when measuring these quantities in Eddington units.
Then we have considered the common dependence upon the redshifts of both luminosities,
and applied the partial Kendall correlation analysis as described in Akritas \& Siebert (1996).
The correlation is still significant, although with a smaller $\tau$.
We then included the upper limits, and repeating the same analysis
we verify that the value of $\tau$ is now greater.

Finally, we applied a simple least square fit and performed a partial correlation analysis
(see Eq. 1 of Padovani 1992), accounting for
the common dependence on the redshift and on the black hole mass of the plotted quantities.
The results are listed in Tab. \ref{correlation}.
In this case we have excluded all upper limits from the analysis.

Before discussing the implications of this correlation,
there are a few caveats to remind, concerning possible important selection effects:
\begin{itemize}

\item 
In the 1LAC catalog there are many detected sources without a known redshift.
As discussed in Abdo et al. (2010b) and in G11, if these sources will turn out to be 
at $z\sim2$, then their $\gamma$--ray luminosities would be huge.
If the absence of broad emission lines is due to their intrinsic weakness, then 
these blazars would be located in the bottom right part of Fig. \ref{class}, and 
they would be clear outliers of the found correlation.
If, instead, the absence of lines is due to a particularly strong non thermal
continuum, then $L_{\rm BLR}$ could be large, locating these objects in the
``FSRQs quadrant".

\item
We have clear examples of blazars varying their $\gamma$--ray luminosity 
by more than 2 orders of magnitude.
It is very likely that the present samples of $\gamma$--ray detected blazars
preferentially include objects in their high state\footnote{This would
also explain why the radio and $\gamma$--ray fluxes are correlated, 
even if only a relatively small fraction of radio--loud AGN with a flat spectrum
are detected in $\gamma$--rays; see e.g. Ghirlanda et al. (2011).}.
The $\gamma$--ray luminosity we have considered is the average over 11 months
(therefore the short term variability is averaged for),
but blazars can indeed be variable over longer periods.
This variability introduces an inevitable dispersion around the
correlation line.

\item
Misaligned jets should be weaker $\gamma$--ray sources than their aligned
counterpart, but they would show the same emission line luminosities.
Therefore weak $\gamma$--ray sources {\it must} exist, populating the region
to the left of the interpolating line of Fig. \ref{class}.
However, these sources would be classified as radio--galaxies
(see Abdo et al. 2010c), and not aligned blazars, although
some overlap might exist.


\end{itemize}

Despite the presence of the above caveats, the apparent correlation between
the BLR and the jet $\gamma$--ray luminosity is certainly intriguing, since 
it would prove
the importance of the emission line photons in the
production of high energy $\gamma$--rays and, more importantly,
it would point
towards a relation between the accretion rate and the jet power.
This relation is not direct, however, since the observed $\gamma$--ray
luminosity can be considered a rather poor proxy of the jet power
and the disc luminosity is linearly related to the accretion rate only
for ``standard" optically thick geometrically thin accretion disc.
This will be discussed more in the next section.

\vskip 0.3 cm

Fig. \ref{class} shows that BL Lac objects are rather neatly divided from FSRQs:
with few exceptions, all FSRQs have $L_{\rm BLR}/L_{\rm Edd}>5\times 10^{-4}$ and
all  BL Lacs are below this value.
The corresponding dividing $\gamma$--ray luminosity is $L_\gamma/L_{\rm Edd}\sim 0.1$.
We derived this apparent ``divide"  considering only the sources 
for which we have a detection of the BLR and the $\gamma$--ray luminosities.
In other words, we first excluded the upper limits from the analysis.
The paucity of data does not allow us to reach a firm conclusion about the 
exact value of this divide, but, reassuringly, when we include all the upper
limits they lie in the ``correct" quadrant of the plane.

On the other hand, most upper limits correspond to BL Lacs with $z<0.4$, 
so an issue remains: the divide between BL Lacs and FSRQs 
could be partly due to a segregation in redshift, if all BL Lacs
are at low redshift and FSRQs at high redshift.
To verify this, we plot in Fig. \ref{blrz} the BLR luminosity (in Eddington units) as
a function of redshift.
It can be seen that there are {\it detected} BL Lacs with $L_{\rm BLR}/L_{\rm Edd} < 5\times 10^{-4}$
at relatively large redshifts.
This is a hint that the divide is real, but, again, the paucity of points
precludes a firmer conclusion about this possible selection effect.
Moreover, most of the upper limits come from the P11 sample, that by construction 
selects only BL Lacs at $z<0.4$.
This limit in redshift possibly introduces a bias in the dividing value,
as can be seen in Fig \ref{blrz}.
Nevertheless, we reiterate that the upper limits did not take part 
in the determination of the divide, hence this bias does not completely compromise the
result.

If real, the found divide would be 
in agreement with what found studying the distribution of bright {\it Fermi}
detected blazars in the $\gamma$--ray spectral index -- $\gamma$--ray luminosity
plane (for them Ghisellini Maraschi \& Tavecchio 2009 proposed a ``divide" between
BL Lacs and FSRQs around $L_{\rm d}/L_{\rm Edd}\sim 10^{-2}$), and to what
more recently found by G11 using a sample of bright blazars much more limited in number
than what we use here.
The value of the divide found here would also be consistent with 
the division between FR I and FR II radio--galaxies, found with a completely
different approach by Ghisellini \& Celotti (2001).

\vskip 0.3 cm

Fig. \ref{class} shows also that all blazars form a continuous family,
with no apparent ``discontinuity" (or sign of bimodality). 
Excluding upper limits, we find that $L_{\rm BLR}\propto L_\gamma$ (normalizing or not
to Eddington).
One can then wonder if it is still meaningful to divide BL Lac objects
from FSRQs, since in Fig. \ref{class} they form a continuous distribution.
In other words: are BL Lacs and FSRQs characterized by some different fundamental
properties, or are they simply at two sides of a continuous distribution
of properties?
An example can illustrate this point:
suppose, as suggested by Ghisellini, Maraschi \& Tavecchio (2009) that the accretion
disc in BL Lacs is radiatively inefficient, while it is efficient in FSRQs.
This is a fundamental different property, although it concerns the accretion disc,
not the jet.
Another example: suppose that jets in BL Lacs are made by pure electron--positron
plasmas, while the jets in FSRQs are made by normal electron--proton plasmas.
This, too, should be considered a fundamental different property.
If instead all blazars have radiatively efficient accretion disc and their jets
are all made by electrons and protons, then they {\it look} different only because
they have different overall powers, and this in turn might also explain
why their SED is different, without the need of anything fundamental dividing them.

The discussion below is dedicated to this issue, focussing in particular to
the proposed ``divide" of BL Lacs and FSRQs in terms of the mass accretion rate
in Eddington units.

\begin{figure}
\vskip -0.6 cm 
\hskip -0.2 cm
\psfig{figure=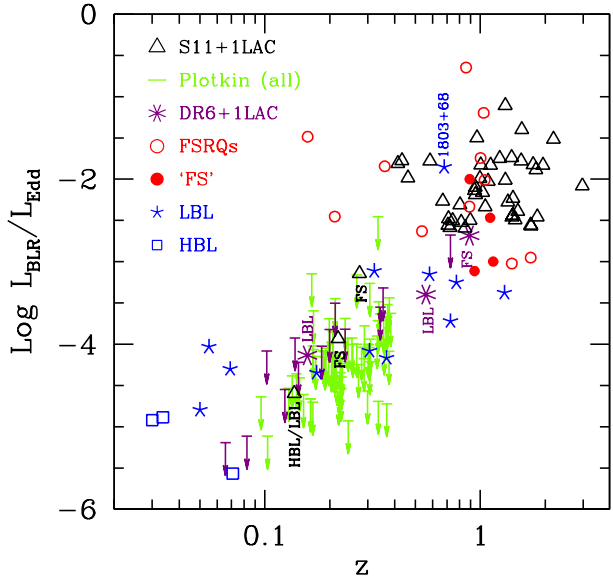,width=9cm,height=9cm}
\vskip -0.6 cm 
\caption{The broad line luminosity (in Eddington units) as a function 
of redshift. Same symbols as in Fig. \ref{class}. 
}
\label{blrz}
\end{figure}

\section{Discussion}

Our study concerns $L_{\rm BLR}$, $L_\gamma$, and the black hole mass.
In the following we will discuss how we can use $L_{\rm BLR}$ to find the
disc luminosity $L_{\rm d}$, and how to use $L_\gamma$ to find a proxy
for the jet power $P_{\rm jet}$ and the mass accretion rate $\dot M$.
The black hole mass is of course used to normalize all powers to the Eddington
luminosity.

\vskip 0.3 cm

$L_{\rm BLR} \to L_{\rm d} \to \dot M$ --- 
For radiatively efficient accretion discs, 
the BLR luminosity is a direct measure of the disc luminosity $L_{\rm d}$,
since, on average, $L_{\rm d}\sim 10\, L_{\rm BLR}$
(see e.g. Baldwin \& Netzer 1978;  Smith et al. 1981). 
Radiatively efficient (i.e. Shakura--Sunjaev 1973) disc
should occur for $\dot m \equiv \dot M/\dot M_{\rm Edd} > \dot m_{\rm c}$.
Defining $\dot M_{\rm Edd}\equiv L_{\rm Edd}/c^2$ (without the efficiency factor), 
then $\dot m_{\rm c}$ should be close to 0.1 (Narayan \& Yi, 1995).
Another hypothesis suggests the lower value $\dot m_{\rm c}\sim10^{-4}$ 
for the radiative transition (Sharma et al. 2007).
If the disc emits as a black--body at all radii, then most of the power 
is emitted in the far UV, and we can approximate the photo--ionizing luminosity
with the entire $L_{\rm d}$.

When $\dot m < \dot m_{\rm c}$, the disc should become radiatively inefficient,
because the particle density of the accretion flow becomes small,
and the energy exchange timescale between protons and electrons becomes
smaller than the accretion time.
If this occurs, the disc bolometric luminosity decreases.
Narayan, Garcia \& McClintock (1997) proposed that in this regime $L_{\rm d} \propto \dot M^2$.
In this case the disc becomes hot, inflates, and it does not emit
black--body radiation. 
As a consequence, $L_{\rm ion} \ll L_{\rm d}$.
According to Mahadevan (1997, see his Fig. 1), the decreasing fraction of the 
ionizing luminosity is as important as the decrease of the overall efficiency
$\eta$ (defined as $L_{\rm d}=\eta \dot M c^2$).
In the example shown by Fig. 1 of Mahadevan (1997), $L_{\rm ion}\propto \dot M^{3.5}$.
If this were true, we expect also the broad emission line luminosity
to have the same dependence on $\dot M$ when $\dot M$ goes sub--critical.
Note, however, that the SED calculated by Mahadevan (1997) could 
be largely affected by the presence of extras sources of seed photons for the thermal Componization,
besides the assumed cyclotron--synchrotron emission.
These extra seed photons (for instance coming for some cool part of the disc)
could enhance the UV emission both by enhancing the Compton scattering and
by cooling the hot emitting electrons.
So we regard the $L_{\rm BLR} \propto L_{\rm ion}\propto \dot M^{3.5}$ relation 
as an indication, but without excluding other possibilities.
In practice, for $\dot m < \dot m_{\rm c}$, 
we will consider both $L_{\rm BLR} \propto \dot M^2$
and $L_{\rm BLR} \propto \dot M^{3.5}$.

\begin{figure}
\vskip -0.6 cm 
\hskip -0.2 cm
\psfig{figure=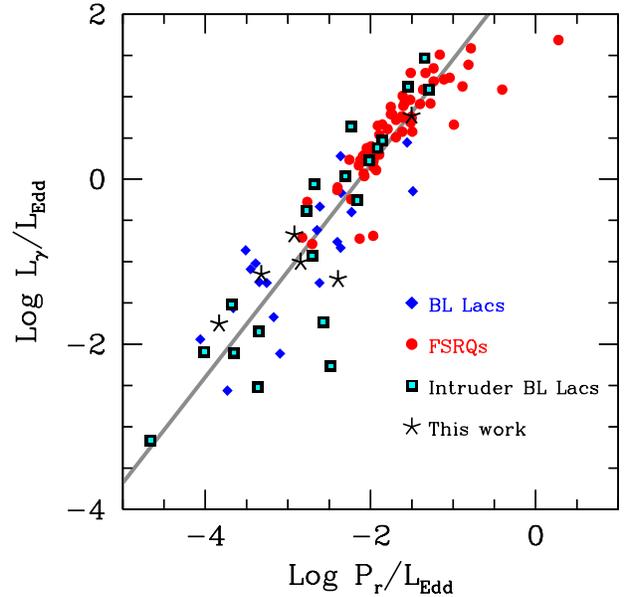,width=9cm,height=9cm}
\vskip -0.4 cm 
\caption{The $\gamma$--ray luminosity in the LAT band as a function of
the jet power $P_{\rm r}$ (both in Eddington units).
The latter is the power that the jet has spent to produce the (bolometric)
radiation we see, and is given by $P_{\rm r}\sim L_{\rm bol}/\Gamma^2$.
To derive it  we have used blazars detected by {\it Fermi} for
which we have constructed the SED and estimated the bulk Lorentz factor
(G10; G11; Tavecchio et al. 2010).
The grey solid line shows the result (bisector) of a least square fit.
}
\label{prlgamma}
\end{figure}

\vskip 0.3 cm
$L_\gamma \to P_{\rm jet} \to \dot M$ ---
A {\it lower limit} on the jet power $P_{\rm jet}$ is
\begin{equation}
P_{\rm jet}\, >\, P_{\rm r}\, \gsim \, {L_{\rm bol} \over \Gamma^2}
\end{equation}
where $\Gamma$ is the bulk Lorentz factor (see Celotti \& Ghisellini 2008 and
Ghisellini \& Tavecchio 2009),
$L_{\rm bol}$ is the jet bolometric luminosity, measured assuming isotropic emission.
$P_{\rm r}$ is the power that the jet has to spend to produce the radiation we see.
If $P_{\rm jet}\sim P_{\rm r}$, then the jet uses its entire power to produce 
radiation, including its kinetic power, and it should stop.
Fitting the SED of a large number of blazars detected by {\it Fermi}
with a simple one--zone leptonic model (as the one used in the Appendix)
returns values of $\Gamma$ contained in a narrow range, around 13--15 
(see Ghisellini et al. 2010; hereafter G10), consistent with the values one 
derives from the superluminal motion.
The same model also yields the number of emitting electrons
required to account for the radiation we see, and by assuming one
proton per electron, it yields $P_{\rm jet} \sim$30--100$ P_{\rm r}$,
on average.
We can conclude that $P_{\rm jet}$ is robustly bound to be larger than $P_{\rm r}
\sim L_{\rm bol}/200$ and it can be a factor 30--100 larger than that.

The key result obtained in G10 (confirming earlier results in Celotti \& Ghisellini 2008 and
then also confirmed by G11), is that $P_{\rm jet} \sim \dot M c^2$.
Since $P_{\rm r}$ is linearly related to $P_{\rm jet}$, and it is a 
model--independent quantity, we can safely use $P_{\rm r}$ as a proxy for
$P_{\rm jet}$ and $\dot M$.
To measure $P_{\rm r} \sim L_{\rm bol}/\Gamma^2 $ we should construct the SED of all our blazars.
On the other hand, we can take advantage from the sample already studied
in the literature, to find if there is a robust relation between $P_{\rm r}$
and the $\gamma$--ray luminosity $L_\gamma$ {\it in the Fermi/LAT energy range},
that is the quantity most readily available to us.
Such a robust relation indeed exists, and it is shown in Fig.  \ref{prlgamma}.
Fitting it  with a simple least square method and taking the bisector we obtain: 
\begin{equation}
{\rm Log} \left( {L_\gamma\over L_{\rm Edd} } \right) =
1.284\; {\rm Log} \left( { P_{\rm r}\over L_{\rm Edd} } \right)   + 2.738
\end{equation}
%
Note that the two quantities are not linearly related.
At low values of $L_\gamma/L_{\rm Edd}$, the $\gamma$--ray luminosity 
{\it underestimate} $P_{\rm r}$
(either normalizing or not to the Eddington luminosity).
As a consequence, the $L_\gamma/L_{\rm Edd}$ values for low luminosity BL Lacs underestimate  
the jet power, and in turn the mass accretion rate.
This occurs because $L_\gamma$ is not  a good indicator of $L_{\rm bol}$ for low
luminosity BL Lacs, that have their high energy SED peaking in the TeV band,
and that have an important (often dominant) synchrotron component.

\vskip 0.3 cm
$L_{\rm BLR}/L_{\rm Edd} $ vs $ P_{\rm r}/L_{\rm Edd}$ ---  
Since the above correlation appears rather robust, we can use it to 
obtain $P_{\rm r}/L_{\rm Edd}$ from our values of $L_\gamma$ and black hole mass, 
without the need to construct the SED and modelling it.
In turn, we can think at the obtained $P_{\rm r}/L_{\rm Edd}$ values as proportional to the
mass accretion rate in Eddington units.
This is done in Fig. \ref{combo} where we plot $L_{\rm BLR}/L_{\rm Edd}$ 
as a function of $P_{\rm r}/L_{\rm Edd}$.

\begin{figure}
\vskip -0.6 cm 
\hskip -1.8 cm
\psfig{figure=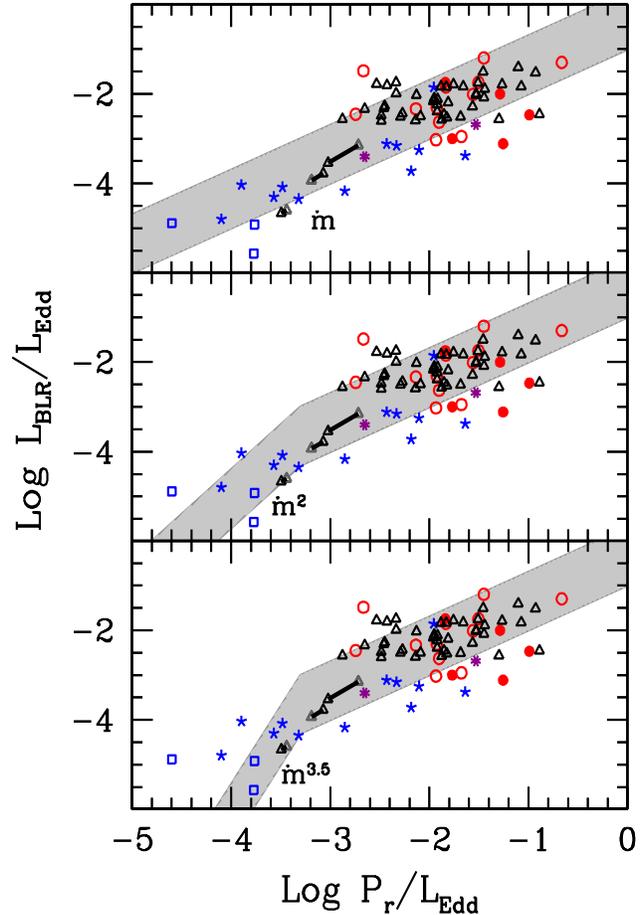,width=12cm,height=13cm}
\vskip -0.3 cm 
\caption{
Luminosity of the broad line region (in Eddington units) for the 
sources from our samples and from G11 as a function of $P_{\rm r}$ 
(in Eddington units).
Symbols are the same as Fig. \ref{class}, but without upper limits and
with grey stripes superimposed.
In all three panels we assume that $P_{\rm r}$ tracks $\dot M$.
If this is true, we can see how the broad line luminosity
is related to $\dot m \equiv \dot M /\dot M_{\rm Edd}$, or, equivalently, to 
$P_{\rm r}/L_{\rm Edd}$.
At large values of $L_{\rm BLR}/L_{\rm Edd}$ the grey stripe
of all panels corresponds to  
$L_{\rm BLR}/L_{\rm Edd}\propto P_{\rm r}/L_{\rm Edd}$.
In the top panel the grey stripe continues to show a linear relation 
also at low values of $L_{\rm BLR}/L_{\rm Edd}$,
while in the mid panel the grey stripe becomes quadratic below some critical value
(here $L_\gamma/L_{\rm Edd}=0.1$ is assumed, namely the value dividing BL Lacs
from FSRQs).
In the bottom panel, the grey stripe becomes 
$L_{\rm BLR}/L_{\rm Edd} \propto (P_{\rm r}/L_{\rm Edd})^{3.5}$
at low values, to account for the deficit of ionizing UV photons
in radiatively inefficient disc.
}
\label{combo}
\end{figure}

For $\dot m > \dot m_{\rm c}$, we expect that the BLR luminosity, in Eddington units,
depends linearly on the normalized accretion rate, and therefore
$L_{\rm BLR}/L_{\rm Edd} \propto \dot m \propto P_{\rm r}/L_{\rm Edd}$. 

Below $\dot m\sim \dot m_{\rm c}$, we expect that the disc becomes radiatively inefficient.
The disc {\it bolometric} luminosity becomes proportional
to $\dot m^2$, while the {\it ionizing} luminosity can follow an even steeper
relation with $\dot m$ (i.e. $\propto \dot m^{3.5}$, see above).
The grey stripe at small $L_{\rm BLR}/L_{\rm Edd}$ is  proportional to  
$P_{\rm r}/L_{\rm Edd}$ in the top panel,  to $(P_{\rm r}/L_{\rm Edd})^2$
in the mid panel and to $(P_{\rm r}/L_{\rm Edd})^{3.5}$ in the bottom panel.
We mainly consider the Narayan \& Yi hypothesis of $m_{\rm c}\sim0.1$.
The break value has been chosen to correspond to $L_{\rm BLR}/L_{\rm Edd} \sim 5\times 10^{-4}$,
and allowing for considerable scatter around this value.
Note that, if we take into account the $m_{\rm c}\sim10^{-4}$ hypothesis 
(Sharma et al.\ 2007), 
the break values would be outside the luminosity range of our sample.

Fig. \ref{combo} shows that we cannot yet distinguish among the different
cases, even if there is some preference for the 
$L_{\rm BLR}/L_{\rm Edd}\propto (P_{\rm r}/L_{\rm Edd})^2$     
case. 

There are BL Lacs of the HBL type, such as Mkn 421, Mkn 501 and 2005--489, that
do show broad emission lines
(and the prototypical BL Lac object too, i.e. BL Lac itself), 
with $L_{\rm BLR}$ around $10^{42}$ erg s$^{-1}$ and $L_{\rm BLR}/L_{\rm Edd}$ around
$10^{-5}$.
These are the objects at the low (bottom--left) end of Fig. \ref{class}, Fig. \ref{prlgamma}
and Fig. \ref{combo}.
We need more low power BL Lacs to investigate if they are
indeed associated with radiatively inefficient discs.
 
{\it Note that this is not required} by the basic explanation of the blazar sequence and
by the proposed luminosity ``divide" between BL Lacs and FSRQs.
In fact, the change of the observed SED along the blazar sequence, interpreted as
a radiatively cooling sequence (Ghisellini et al. 1998), requires that 
in BL Lacs the emission lines are not important as seed photons for the inverse Compton
process.
This can be the case {\it even if the lines are present}, if the dissipation region
occurs outside $R_{\rm BLR}$: in this case the BLR photons are seen in the comoving region
red--shifted and time--diluted, and the EC process can be negligible.
The relation between the size of the BLR and the ionizing luminosity ensures that 
in BL Lacs the size of the BLR is much smaller than in FSRQs, even if the black hole
mass is similar.
If dissipation occurs always at $R_{\rm diss} \sim 10^3$ \sc\ radii, then in BL Lacs
we easily have $R_{\rm diss}>R_{\rm BLR}$ (and emitting lines are negligible for
the formation of the high energy spectrum), while in more powerful FSRQs 
we have $R_{\rm diss}<R_{\rm BLR}$, with a corresponding enhancement of the EC process.
Earlier works (Celotti \& Ghisellini 2008; G10, G11) have shown that $R_{\rm diss}$
is of the order of $\sim 10^3$ \sc\ radii in all objects.
Requiring that the size of the BLR is a factor $f$ smaller than this,
and using $R_{\rm BLR}=10^{17} L_{\rm d, 45}^{1/2}$ cm,
we obtain
\begin{equation}
R_{\rm BLR} \, < f R_{\rm diss} \, \to \, 
{L_{\rm d}\over L_{\rm Edd}} < 6.9\times 10^{-3} f^2 M_8 \left( {R_{\rm diss}\over 10^3\, R_{\rm S} } \right)^2
\end{equation}
where $M=10^8 M_8$ solar masses.
We obtain, in this case, a value for the divide in agremeent with 
the observed $L_{\rm BLR}/L_{\rm Edd}\sim 5\times 10^{-4}$ (for $L_{\rm BLR}\sim 0.1 L_{\rm d}$ and
$f$ smaller than, but close to unity), but dependent on the black hole mass.
The dependence on the black hole mass would produce some blur in the division,
that is not inconsistent with what we see.




\section{Conclusions}

In this work we have studied those blazars detected by {\it Fermi}/LAT
and present in the SDSS optical survey, for which the redshift is known
and there is a black hole mass estimate.
From the broad emission line luminosities (or their upper limits)
we have calculated the luminosity of the entire Broad Line Region,
used as a proxy for the luminosity of the accretions disc.
We could find values for both the BLR and the $\gamma$--ray luminosity
for 78 
blazars, values for $L_{\rm BLR}$ and upper limits on $L_\gamma$ 
for 23 blazars, and upper limits on both the quantities for 62 sources.
Our results can be summarized as follows:
\begin{enumerate}
\item
The luminosity of the BLR correlates well with the $\gamma$--ray luminosity
in the {\it Fermi}/LAT energy range. The correlation is linear, irrespective
if the above luminosities are normalized to Eddington or not.
All upper limits (not used to find the correlation) are consistent with the
correlation itself.

\item
BL Lac objects and FSRQs occupy different regions of the $L_{\rm BLR}/L_{\rm Edd}$--$L_\gamma/L_{\rm Edd}$
plane, with a division at about $L_{\rm BLR}/L_{\rm Edd}\sim 5\times 10^{-4}$.
This confirms, with an enlarged sample, earlier results.
Nevertheless, since the sample is still rather poor of sources with detections 
on both $L_{\rm BLR}$ and $L_\gamma$ and instead rich of upper limits, 
this ``divide" still needs further studies with a more populated sample.

\item
For objects (analyzed in previous works) of known $L_\gamma$, $P_{\rm r}$,
and black hole mass, there is a strong correlation between the two quantities, both using
absolute values and normalizing them to the Eddington luminosity:
$(L_\gamma / L_{\rm Edd}) \propto (P_{\rm r} / L_{\rm Edd})^{1.28}$.
As a consequence, the $\gamma$--ray luminosity (in the {\it Fermi}/LAT
energy range) can be used to estimate $P_{\rm r}$ which is a robust proxy for the 
jet power.

\item
The relation between the strength of the emission lines and the
accretion rate can be used to test radiatively inefficient disc models
and the prediction about the production, in these discs, of the
ionizing luminosity.
Our results are too primitive to draw strong conclusions, but 
there is the possibility that at low accretion rates the 
produced ionizing UV luminosity is larger than expected.

\item
The division between BL Lacs and FSRQs could be due to the transition
between a radiatively inefficient disc to a standard (Shakura--Sunyaev) disc.
Alternatively, it can be due to the dissipation region of the jet being located outside
or inside the BLR.

\end{enumerate}

\section*{Acknowledgments}
We thank the referee for useful comments that improved the paper.
We thank Julian Krolik for suggesting us the possibility of a transition
to ADAF at accretion rates smaller than what we considered.
We also thank F. Tavecchio, G. Ghirlanda and L. Foschini for discussions.
This research made use of the NASA/IPAC Extragalactic Database (NED) 
which is operated by the Jet Propulsion Laboratory, Caltech, under contract 
with NASA, and of the {\it Swift} public data
made available by the HEASARC archive system.

\newpage

\section{Appendix}

\subsection{Spectral energy distribution}
We have characterized the Spectral Energy Distribution (SED)
of the  six sources for which we have both the spectroscopic optical data
and the detection by {\it Fermi}.

To this aim we have collected the data from the NASA Extragalactic Database (NED)
and including the LBAS and 1LAC {\it Fermi}/LAT data (Abdo et al. 2010a; 2010b).

\subsection{The model}
\label{model}

To model the SED we have used the leptonic,
one--zone synchrotron and inverse Compton model, 
fully discussed in Ghisellini \& Tavecchio (2009).

In brief, we assume that in a spherical region of radius $R$, located at a distance
$R_{\rm diss}$ from the central black hole, relativistic electrons are injected at
a rate $Q(\gamma)$ [cm$^{-3}$ s$^{-1}$] for a finite time equal to the 
light crossing time $R/c$. 
For the shape of $Q(\gamma)$ we adopt a smoothly broken power law,
with a break at $\gamma_{\rm b}$:
\begin{equation}
Q(\gamma)  \, = \, Q_0\, { (\gamma/\gamma_{\rm b})^{-s_1} \over 1+
(\gamma/\gamma_{\rm b})^{-s_1+s_2} }
\label{qgamma}
\end{equation}
The emitting region is moving with a  velocity $\beta c$
corresponding to a bulk Lorentz factor $\Gamma$.
We observe the source at the viewing angle $\theta_{\rm v}$ and the Doppler
factor is $\delta = 1/[\Gamma(1-\beta\cos\theta_{\rm v})]$.
The magnetic field $B$ is tangled and uniform throughout the emitting region.
We take into account several sources of radiation externally to the jet:
i) the broad line photons, assumed to re--emit 10\% of the accretion luminosity
from a shell--like distribution of clouds located at a distance 
$R_{\rm BLR}=10^{17}L_{\rm d, 45}^{1/2}$ cm;
ii) the IR emission from a dusty torus, located at a distance
$R_{\rm IR}=2.5\times 10^{18}L_{\rm d, 45}^{1/2}$ cm;
iii) the direct emission from the accretion disc, including its X--ray corona;
iv) the starlight contribution from the inner region of the host galaxy;
v) the cosmic background radiation.
All these contributions are evaluated in the blob comoving frame, where we calculate the 
corresponding inverse Compton radiation from all these contributions, and then transform
into the observer frame.
The latter two contributions are negligible for our sources.

We calculate the energy distribution $N(\gamma)$ [cm$^{-3}$]
of the emitting particles at the particular time $R/c$, when the injection process ends. 
Our numerical code solves the continuity equation which includes injection, 
radiative cooling and $e^\pm$ pair production and reprocessing. 
Our is not a time dependent code: we give a ``snapshot" of the 
predicted SED at the time $R/c$, when the particle distribution $N(\gamma)$ 
and consequently the produced flux are at their maximum.

To calculate the flux produced by the accretion disc, we adopt a standard 
Shakura \& Sunyaev (1973) disc (see Ghisellini \& Tavecchio 2009).

The resulting SEDs and models are shown in Fig. \ref{sed1}, Fig. \ref{sed2} and Fig. \ref{sed3},
and the model parameters are reported in Tab. \ref{para} and Tab. \ref{powers}.

\subsection{Specific sources}

We here discuss briefly a few objects for which we do have both 
the information on the BLR and the $\gamma$--ray luminosity.
These sources lye close to the ``intermediate zone" between BL Lacs and FSRQs.
We have constructed their SED and modeled it through a simple one--zone
leptonic model, as described in Section \ref{model}, in order to classify them. 
To this aim we adopt the same SED--based classification scheme 
discussed in G11 
and originally introduced by Padovani \& Giommi (1995).
In brief, we can classify the object as a FSRQ if the $\gamma$--ray luminosity is
dominating the electromagnetic output and if the X--ray spectrum is flat
(X--ray energy spectral index $\alpha_x<1$); it is a Low frequency peaked BL Lac
(LBL) if the $\gamma$--ray luminosity is comparable to the synchrotron one
and if $\alpha_x<1$, and high frequency peaked BL Lac (HBL) if the 
$\gamma$--ray luminosity is comparable or less than the synchrotron one and
$\alpha_x>1$. 
Fig. \ref{sed1}, Fig. \ref{sed2} and Fig. \ref{sed3} 
show the SED of these sources.
We here summarize our finding.

\vskip 0.2 cm
\noindent
{\bf 0937+5008 ---}  This source is included in the S11 catalog and it 
is detected by {\it Fermi}.
Looking at the SED it can be clearly classified as a FSRQ,
as suggested also from the evident broad H$\alpha$ and H$\beta$ lines 
visible in the SDSS spectrum. From the SED modeling, 
a synchrotron contamination of the optical continuum is visible.
Therefore, the automatic virial black hole mass estimate performed by S11
can be imprecise.
Hence, we chose to assign to its black hole an average mass value
$M=5\times 10^8 M_\odot$.

\vskip 0.2 cm
\noindent
{\bf 1040+24 ---}  This is a DR6 \textit{Fermi} detected source, 
hence the black hole mass is not measured. 
We can assume an average value of $M=5\times 10^8 M_\odot$.
In the SDSS spectrum, a broad MgII line is clearly visible. From 
the SED modeling, this source can be classified as a LBL.
This means that the thermal continuum can be highly contaminated by the
synchrotron emissions, although some broad emission 
lines clearly emerge.
Moreover, the disc seems to be only partially covered, and the synchrotron 
component looks really variable.
This could result in a variable line EW.

\vskip 0.2 cm
\noindent
{\bf 1055+01 ---} This source belongs to the DR6+1LAC sample, 
hence the mass is not measured and we assume the average value
$M=5\times 10^8 M_\odot$.
As in the case of 1040+24, a broad MgII line is visible, but it
is narrower than the usual broad emission line width 
(FWHM$\simeq2500$ km s$^{-1}$). 
This might suggest a small black hole mass, or maybe the line is partially 
covered by the continuum and hence the measure is uncertain.
The SED, indeed, shows that the accretion disc contribution 
is mostly covered by the synchrotron emission.
Overall, we can classify this source as a FSRQ, even if the disc
emission is dominated by the synchrotron radiation.

\vskip 0.2 cm
\noindent
{\bf 1106+023 ---} 
It is a DR6 source detected by \textit{Fermi}.
In this case we can estimate its mass with the Chiaberge
\& Marconi (2011) relation.
In fact the FWHM of its broad H$\beta$ line is reported in the DR6 catalog.
The mass estimate that we obtain is very small ($M=4\times 10^7 M_\odot$). From 
the SED, we can see that the disc emerges from the synchrotron component,
hence the emission lines can be seen clearly. From 
the SED we can classify this source as a LBL.

\vskip 0.2 cm
\noindent
{\bf 1117+2014 ---} This source is present in the S11 catalog,
hence it is supposed to be a quasar.
However, in the SDSS spectrum, the H$\beta$ line is really faint.
Indeed, from the SED modeling, we can classify this source as an HBL.
The SED shows clearly that the synchrotron component largely dominates the 
disc emission, hence the automatic mass estimate performed by S11
($M=4\times 10^8 M_\odot$) is not accurate.
We then assume for this source a value of the black hole of 
equal to the average value.

\vskip 0.2 cm
\noindent
{\bf 1509+022 ---} This is a source included in the S11 catalog.
Looking at the SED, we can classify it as a FSRQ.
However, also in this case the continuum appears contaminated
by the synchrotron component.
Moreover, the S11 results about this source are quite unclear.
The equivalent widths reported in S11 (EW$\sim25-27$ \AA), 
do not seem to be recognizable in the spectrum
(the lines are hardly visible).
Therefore, we think that the S11 black hole mass estimate
can be considered imprecise, and we replace it with the 
average value $M=5\times 10^8 M_\odot$.

\begin{figure}
\vskip -0.6 cm  
\psfig{figure=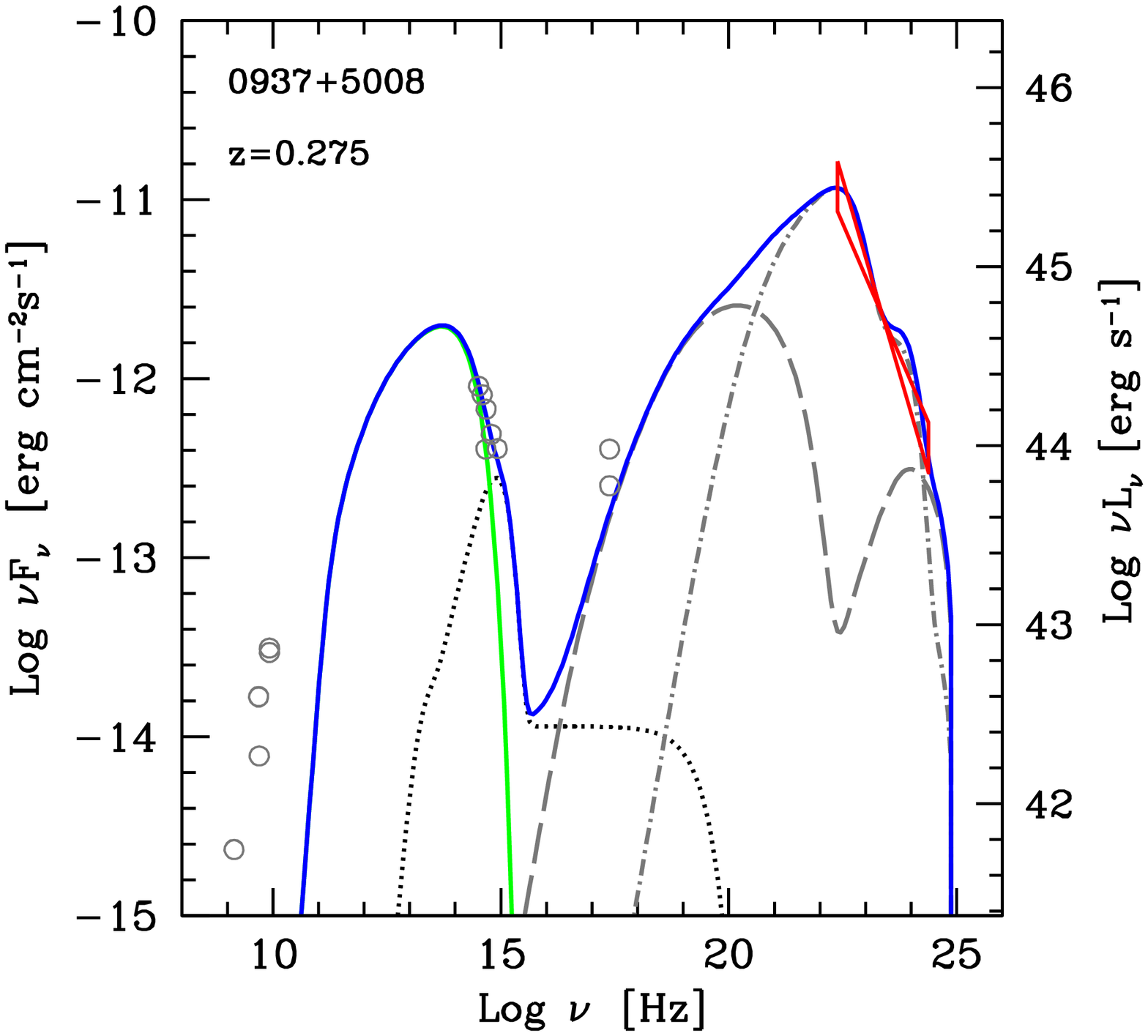,width=9cm,height=8cm}
\vskip -0.8 cm
\psfig{figure=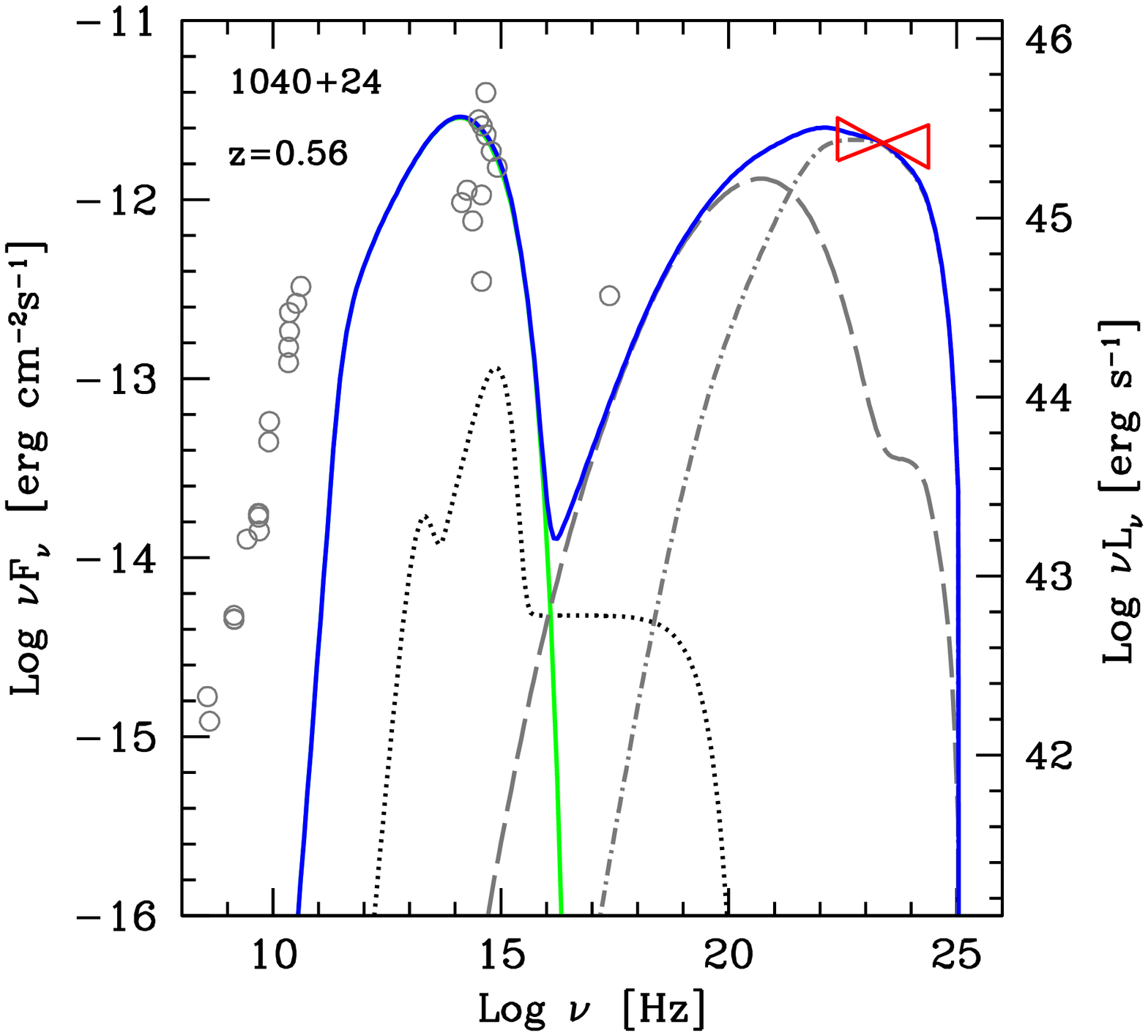,width=9cm,height=8cm}
\vskip -0.8 cm
\caption{SED of 0937+5508 and 1040+24 and the fitting model.
The dotted line corresponds to emission from the accretion disc, the IR torus
(if present) and the X--ray corona.
the thin (green) solid line is the synchrotron component,
the long dashed line the SSC emission, the dot--dashed line the EC 
contribution.
The thick (blue) solid line) is the sum.
}
\label{sed1}
\end{figure}

\begin{figure}
\vskip -0.6 cm 
\psfig{figure=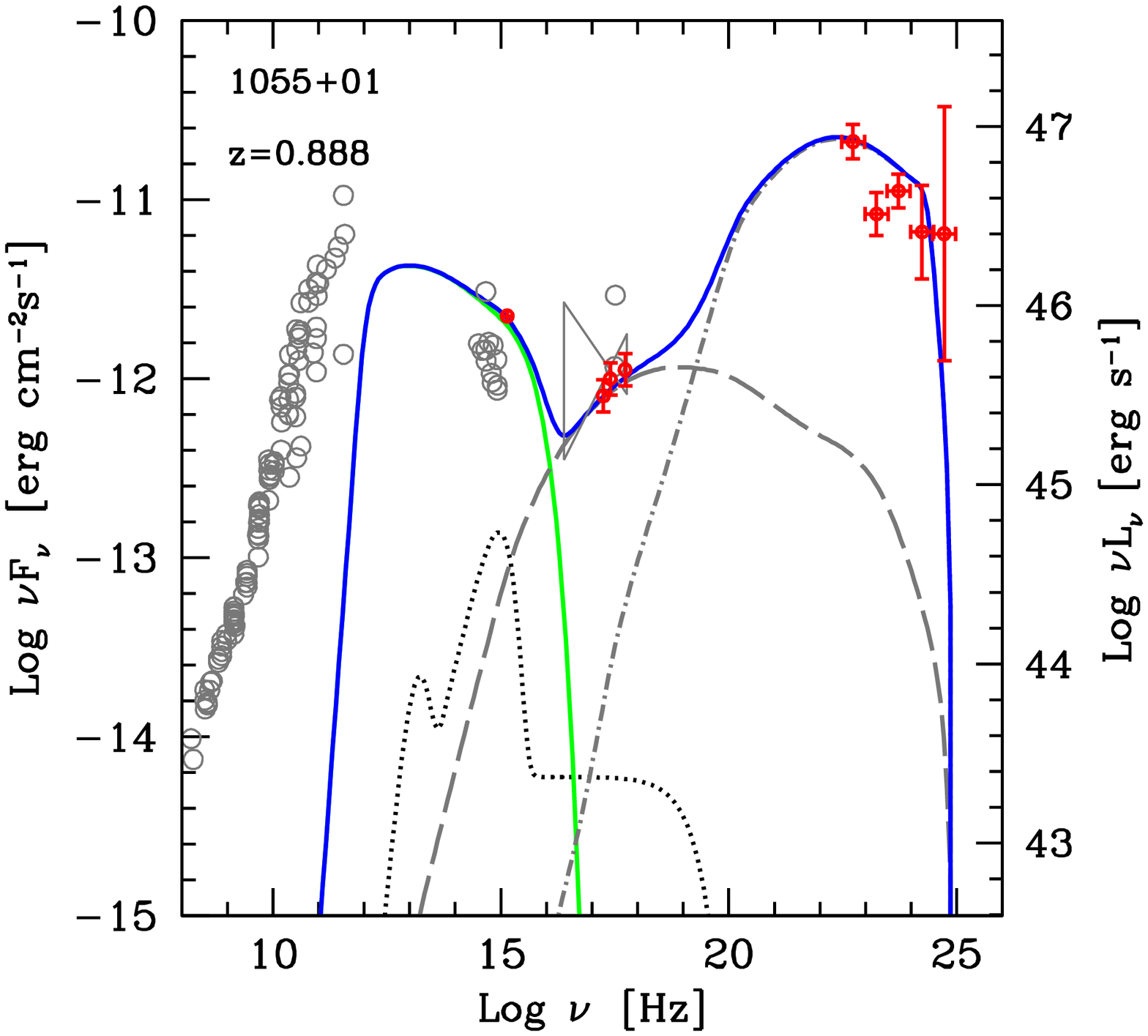,width=9cm,height=8cm}
\vskip -0.8 cm
\psfig{figure=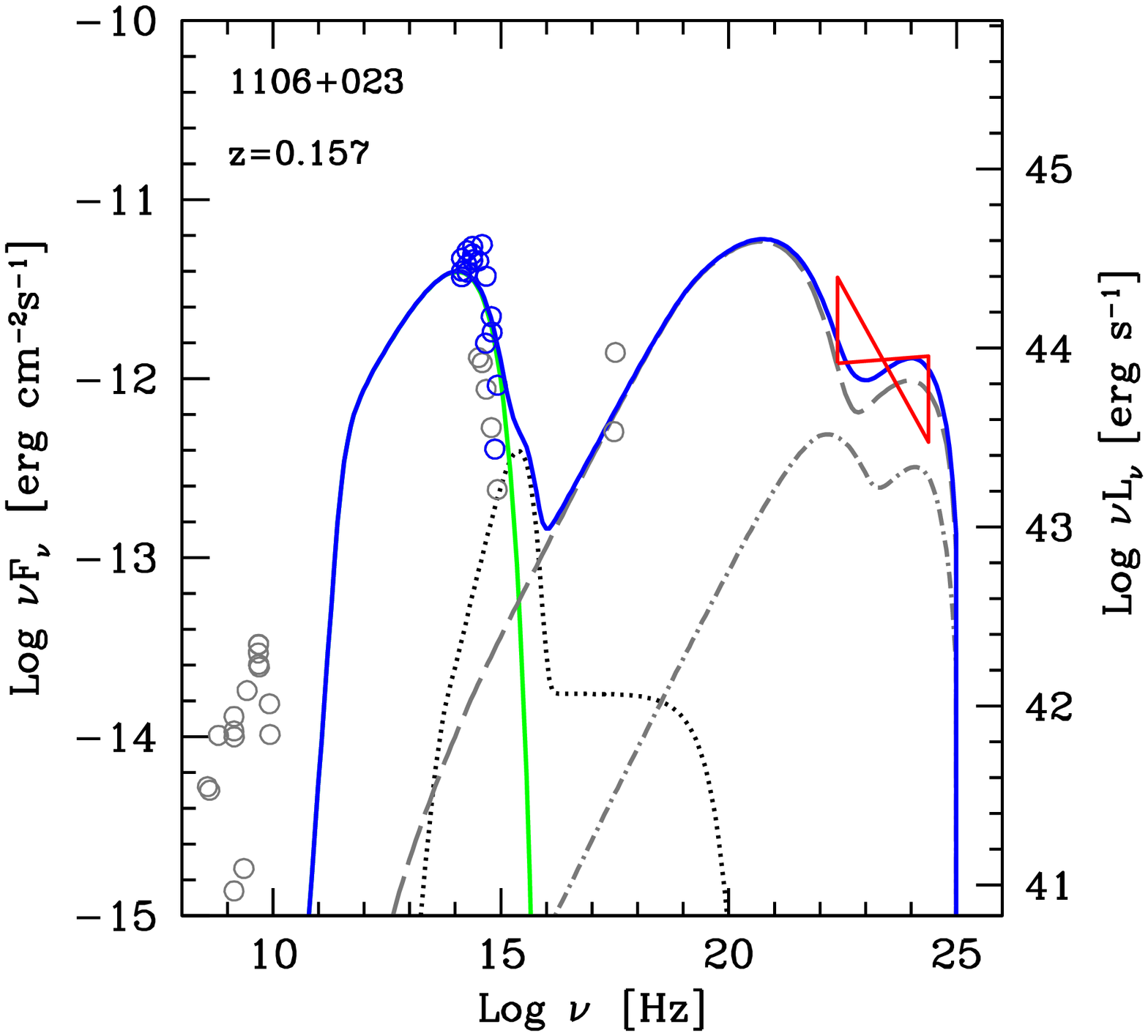,width=9cm,height=8cm}
\vskip -0.8 cm
\caption{SED of 1055+01 and 1106+023. Lines as in Fig. \ref{sed1}.
}
\label{sed2}
\end{figure}

\begin{figure}
\vskip -0.6 cm 
\psfig{figure=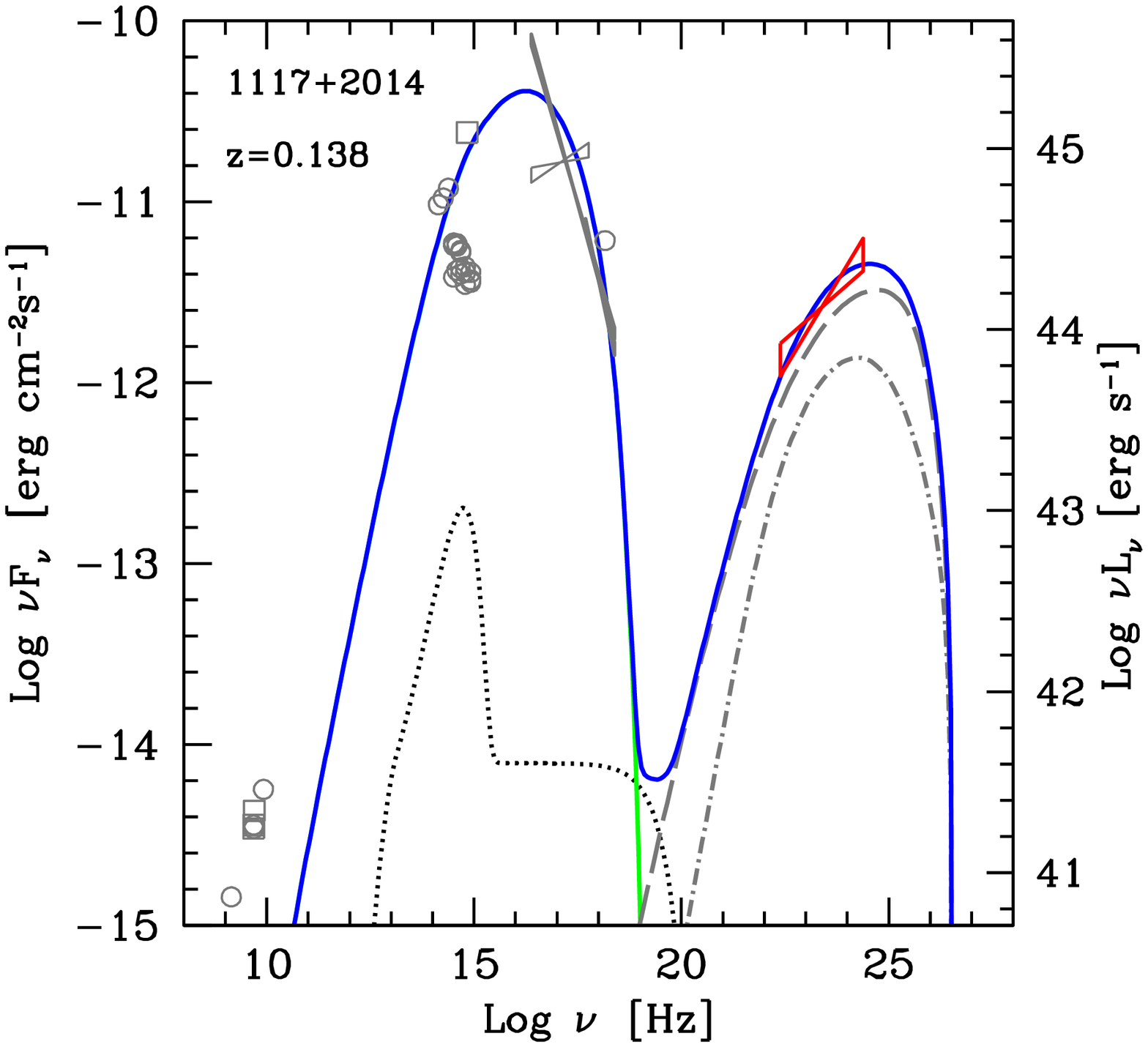,width=9cm,height=8cm}
\vskip -0.8 cm
\psfig{figure=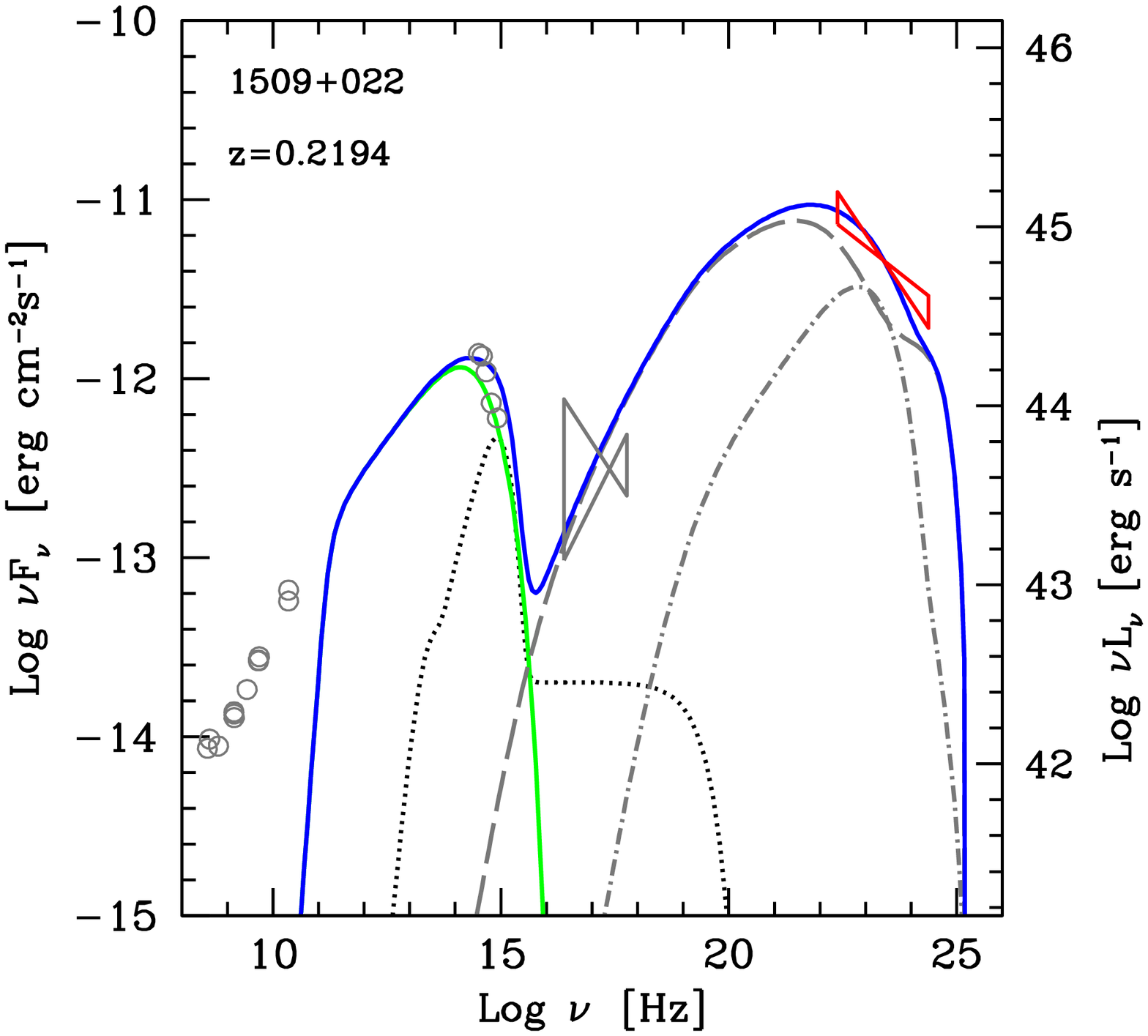,width=9cm,height=8cm}
\vskip -0.8 cm
\caption{SED of 1117+2014 and 1509+022. Lines as in Fig. \ref{sed1}.
}
\label{sed3}
\end{figure}

\begin{table*} 
\centering
\begin{tabular}{llllllllllllll}
\hline
\hline
Name   &$z$ &$R_{\rm diss}$ &$M$ &$R_{\rm BLR}$ &$P^\prime_{\rm i}$ &$L_{\rm d}$ &$B$ &$\Gamma$ &$\theta_{\rm v}$
    &$\gamma_{\rm b}$ &$\gamma_{\rm max}$ &$s_1$  &$s_2$  \\
~[1]      &[2] &[3] &[4] &[5] &[6] &[7] &[8] &[9] &[10] &[11] &[12] &[13] &[14] \\
\hline   
0937+5008  &0.275  &90   (600)   &5e8$^*$ &34.6 &2e--3    &0.12  (1.6e--3) &0.14 &14 &3   &400 &5e3   &0   &2.5  \\ 
1040+23    &0.56   &97.5 (650)   &5e8$^*$ &51.2 &1.5e--3  &0.26  (3.5e--3) &0.67 &13 &3   &100 &9e3   &0   &2 \\ 
1055+01    &0.888  &105  (700)   &5e8$^*$ &98.7 &0.02     &0.98  (0.013)   &3.1  &11 &3.7 &400 &9e3   &1   &2.5 \\ 
1106+023   &0.157  &43.2 (3.6e3) &4e7     &21.9 &3e--3    &0.048 (8e--3)   &0.22 &13 &3   &6e3 &6e3   &2   &2  \\   
1117+2014  &0.138  &105  (700)   &5e8$^*$ &13.7 &5.e--5   &0.019 (2.5e--4) &0.7  &15 &2   &3e4 &1.5e5 &0.5 &3 \\   
1509+022   &0.2194 &120  (800)   &5e8$^*$ &34.6 &0.022    &0.12  (1.6e--3) &0.13 &11 &6   &100 &2e4   &0   &2.3 \\ 
\hline
\hline 
\end{tabular}
\vskip 0.4 true cm
\caption{List of parameters used to construct the theoretical SED.
Col. [1]: name;
Col. [2]: redshift;
Col. [3]: dissipation radius in units of $10^{15}$ cm and (in parenthesis) in units of \sc\ radii;
Col. [4]: black hole mass in solar masses, the asterisk means that the mass is assumed;
Col. [5]: size of the BLR in units of $10^{15}$ cm;
Col. [6]: power injected in the blob calculated in the comoving frame, in units of $10^{45}$ erg s$^{-1}$; 
Col. [7]: accretion disc luminosity in units of $10^{45}$ erg s$^{-1}$ and
        (in parenthesis) in units of $L_{\rm Edd}$;
Col. [8]: magnetic field in Gauss;
Col. [9]: bulk Lorentz factor at $R_{\rm diss}$;
Col. [10] viewing angle $\theta_{\rm v}$ in degrees;
Col. [11] and [12]: break and maximum random Lorentz factors of the injected electrons;
Col. [13] and [14]: slopes of the injected electron distribution [$Q(\gamma)$] below and above $\gamma_{\rm b}$;
The total X--ray corona luminosity is assumed to be in the range 10--30 per cent of $L_{\rm d}$.
Its spectral shape is assumed to be always $\propto \nu^{-1} \exp(-h\nu/150~{\rm keV})$.
}
\label{para}
\end{table*}
%

\begin{table} 
\centering
\begin{tabular}{lllll}
\hline
\hline
Name   &$\log P_{\rm r}$ &$\log P_{\rm B}$ &$\log P_{\rm e}$ &$\log P_{\rm p}$ \\
\hline   
0937+5008 &43.49 &42.06 &44.47 &45.05 \\ 
1040+23   &43.89 &43.43 &44.08 &44.99 \\
1055+01   &45.31 &44.69 &44.68 &46.67 \\ 
1106+023  &42.87 &41.76 &44.58 &46.94 \\
1117+2014 &42.98 &43.66 &42.46 &42.27 \\  
1509+022  &44.42 &42.04 &45.04 &46.03 \\
\hline
\hline 
\end{tabular}
\vskip 0.4 true cm
\caption{
Logarithm of the jet power in the form of radiation, Poynting flux,
bulk motion of electrons and protons (assuming one proton
per emitting electron). Powers are in erg s$^{-1}$.
}
\label{powers}
\end{table}

\end{document}